\documentclass[aps,amsmath,amssymb,twocolumn]{revtex4}

\usepackage{graphicx,color}
 \graphicspath{{./}{./figures/}}
\usepackage{dcolumn}% Align table columns on decimal point
\usepackage{bm}% bold math
\usepackage{feynmp-auto}

\definecolor{labelkey}{cmyk}{.4,.2,0,0}

\newcommand{\blue}{\color{blue}}

\newcommand{\be}{\begin{equation}}
\newcommand{\ee}{\end{equation}}
\newcommand{\bea}{\begin{eqnarray}}
\newcommand{\eea}{\end{eqnarray}}

\newcommand{\nn}{\nonumber }

\newcommand{\JR}{{\mathbb R}}

\begin{document}

\title{Correlations between avalanches in the depinning dynamics of elastic interfaces}
\author{Pierre Le Doussal \textsuperscript{1} and Thimoth\'ee Thiery\textsuperscript{2}}

\affiliation{\textsuperscript{1} 
Laboratoire de Physique de l'\'Ecole Normale Sup\'erieure,\\
ENS, Universit\'e PSL, CNRS, Sorbonne Universit\'e,\\
Universit\'e Paris-Diderot, Sorbonne Paris Cit\'e,\\
24 rue Lhomond, 75005 Paris, France\\
%Laboratoire de Physique Th\'eorique de l'\'Ecole Normale Sup\'erieure,
%PSL University, CNRS, Sorbonne Universit\'es
%24 rue Lhomond, 75231 Paris Cedex 05, France
}
\affiliation{\textsuperscript{2} Instituut voor Theoretische Fysica, KU Leuven, Leuven, Belgium.}
%\date{today}

% \author{Thimoth\'ee Thiery, Pierre Le Doussal} 
% \affiliation{
% Laboratoire de Physique Th\'eorique de l'Ecole Normale Sup\'erieure, PSL University,\\
% CNRS, Sorbonne Universit\'es, 24 rue Lhomond, 75231 Paris Cedex 05, France
% %CNRS-Laboratoire
% %de Physique Th{\'e}orique de l'Ecole Normale Sup{\'e}rieure, 24 rue
% %Lhomond,75231 Cedex 05, Paris, France
% } 
% \date{\today}

\begin{abstract}
We study the correlations between avalanches in the depinning dynamics of elastic interfaces
driven on a random substrate. 
In the mean field theory (the Brownian force model), it is known that the avalanches 
are uncorrelated. Here we obtain a simple field theory which describes the first deviations from this uncorrelated behavior in a $\epsilon=d_c-d$ expansion below the upper critical dimension $d_c$ of the model. 
We apply it to calculate the correlations between (i) avalanche sizes (ii) avalanche dynamics 
in two successive avalanches, or more generally, in two avalanches separated 
by a uniform displacement $W$ of the interface. For (i) we obtain the correlations of the total sizes, of the local sizes and of the total sizes with given seeds (starting points). For (ii) we obtain the correlations of the velocities, of the durations, and of the avalanche shapes. In general we find that the avalanches are {\it anti-correlated},
the occurence of a larger avalanche making more likely the occurence of a smaller one, and vice-versa.
Examining the universality of our results leads us to conjecture several new exact scaling relations for the critical exponents that characterize the different distributions of correlations. The avalanche
size predictions are confronted to numerical simulations for a $d=1$ interface with short range elasticity. They are
also compared to our recent related work on static avalanches (shocks). Finally we show that
the naive extrapolation of our result into the thermally activated creep regime at finite temperature, 
predicts strong positive correlations between the forward motion events, as recently observed
in numerical simulations. 
\end{abstract}

\maketitle

\begin{widetext}

\section{Introduction}

The motion of elastic interfaces slowly driven in a random medium is not smooth but proceeds via jumps extending over a broad range of space and time scale \cite{DSFisher1998,SethnaDahmenMyers2001}.
This avalanche motion is ubiquitous in a number of experimental systems such as magnetic domain walls \cite{ZapperiCizeauDurinStanley1998,DurinZapperi2000}, fluid contact lines  \cite{MoulinetGuthmannRolley2002,LeDoussalWieseMoulinetRolley2009}, earthquakes \cite{BenZionRice1993,FisherDahmenRamanathanBenZion1997}, cracks \cite{Ponson2008,SantucciGrobToussaint2010,BonamySantucciPonson2008} or imbibition fronts \cite{PlanetRamonSantucci2009}, often modeled as elastic interfaces.
Theoretically, the statics and the dynamics of elastic interfaces has been studied using the functional renormalization group (FRG)\cite{DSFisher1986,NattermannStepanowTangLeschhorn1992,NarayanDSFisher1992b,NarayanDSFisher1993a,ChauveLeDoussalWiese2000a,LeDoussalWieseChauve2002,LeDoussalWieseChauve2003,LeDoussal2006b,LeDoussal2008,MiddletonLeDoussalWiese2006,RossoLeDoussalWiese2006a,RossoLeDoussalWiese2009a}. 
The FRG has then been extended to study avalanches, either in the statics (the so-called
shocks) \cite{LeDoussalWiese2008c,LeDoussalWiese2011b}, or near the depinning transition 
\cite{LeDoussalWiese2011a,LeDoussalWiese2012a,DobrinevskiLeDoussalWiese2014a,DobrinevskiPhD,ThieryPhD,Tip,ThieryShape}.

An important question is to quantify the temporal and spatial correlations 
between successive avalanches. It is well known that in the case of earthquakes strong temporal correlations are observed, called aftershocks \cite{Omori1894}. It was believed that in the context of elastic interfaces models, correlations between avalanches arise only if one includes additional mechanisms in the interface dynamics, such as relaxation processes \cite{JaglaLandesRosso2014,Jagla2014} or memory effects \cite{DobrinevskiLeDoussalWiese2013}. In a recent work \cite{ThieryStaticCorrelations}
we have studied correlations between ``static avalanches", more precisely between the sizes and locations of the shocks in the ground state of elastic interfaces in a random potential. Although they are expected to be close cousins of the avalanches observed in the interface dynamics, they are not identical. Thus it remains to study the correlations between 
avalanches in the dynamics. 

In this paper we study the correlations between avalanches in the depinning dynamics of elastic interfaces
driven on a random substrate. The starting point is the mean-field theory, valid in space dimension
$d>d_c$, known as the Brownian force model (BFM)
\cite{LeDoussalWiese2011b,LeDoussalWiese2011a,LeDoussalWiese2012a,ThieryLeDoussalWiese2015,Delorme}
a multidimensional generalization of the celebrated ABBM model \cite{AlessandroBeatriceBertottiMontorsi1990}. In the BFM, the avalanches are strictly uncorrelated \cite{ThieryLeDoussalWiese2015}.
Here we obtain a simple field theory, based on the FRG, which describes the first deviations from this uncorrelated behavior in a $\epsilon=d_c-d$ expansion below the upper critical dimension of the model
(which depends on the range of the elastic interaction, $d_c=4$ for short-range (SR) elasticity and $d_c=2$ for usual long-range (LR) elasticity). The elastic model and the avalanche observables are defined in Section \ref{sec:ModelAndObservables}. The field theory is described in \ref{sec:analytical:fieldtheory} and
Appendix \ref{app:derivation}, 
together with a discussion of the physical origin of the correlations. 

We apply our theory to calculate the correlations of two successive avalanches, loosely meaning two avalanches which occur within the same dynamical forward evolution of the interface in a given pinning landscape. It is convenient to study two avalanches separated by a given displacement $W$ of the center of mass of the interface. For $W=0^+$ this describes immediately successive avalanches. We study two types of information, (i) the correlations of the avalanche sizes, in Section \ref{sec:size}, 
and (ii) the correlations in the dynamics within each avalanche, in Section \ref{sec:dynamical}.
More precisely for (i) we obtain the correlation between the total sizes, the local sizes, and the
total size of avalanches with given seeds (i.e. given position of their starting points).
We show that the first two results are equal to this order in the expansion, i.e. to $O(\epsilon)$, to
the ones obtained in the statics \cite{ThieryStaticCorrelations} for random field disorder
(differences are expected to the next order).
These results are derived here in a much simpler fashion. The third one, the
correlation of the total size as the distance between the seeds is varied, is new. Some of these analytical predictions are confronted, in Section \ref{sec:numerics},
to numerical simulations for a $d=1$ interface with short range elasticity. For the dynamics (ii)
we obtain the correlations of the velocities, of the avalanche durations, and of the avalanche shapes, i.e.
of the velocity as a function of time at fixed duration. For the latter, a deviation from the famous parabola shape is demonstrated in the correlation. Examining the universal limit of our results, we obtain some new non-trivial (and presumably exact, i.e. valid beyond the $\epsilon$ expansion) conjectures for a variety of critical exponents that characterize the correlations.

We find that in the depinning dynamics the avalanches are {\it anti-correlated}, 
the occurence of a larger avalanche (in size, in total velocity etc..) 
making more likely the occurence of a smaller one, and vice-versa. The same was observed in the statics
(i.e for correlations between shocks) for random field disorder, while both positive and negative correlations
could occur for random bond disorder depending on $W$. In the conclusion we discuss qualitatively
some possible extension at finite temperature which indicates instead the occurence
of positive correlations in the creep regime. 

% Our results reproduce the results previously obtained in the statics, obtained here in a much simpler fashion.
% We also extend the calculation to compute the correlations between the sizes
% of the avalanches conditioned on starting at specific points (the seeds). 

\section{Model and observables} \label{sec:ModelAndObservables}

\subsection{Model} \label{subsec:Model}

We focus on a $d$-dimensional elastic interface whose position at point $x \in \JR^d$ and time $t \in \JR$, $u(x , t) \in \JR $, satisfies the following equation of motion \footnote{we use interchangeably $\dot u$ or $\partial_t u$ to
denote partial derivatives w.r.t. time}

\begin{equation} \label{eq:model:intro}
\eta \partial_t u(x,t) = \nabla^2_x u(x,t)- m^2( u(x,t) -w(x,t) ) + F(u(x,t),x) \, .
\end{equation}
The random pinning force $F(u,x)$ is chosen Gaussian with correlator
\begin{equation}
\overline{F(u,x)F(u',x')} = \delta^d (x-x') \Delta_b(u-u') \, .
\end{equation}
where $\Delta_b(u)$ denotes the bare correlator, assumed to be a symmetric short-range function.
Here we have restricted to elastic interfaces with short-range elasticity and the elastic coefficient (coefficient in front of the Laplacian in \eqref{eq:model:intro}) has been set to unity by a choice of units. The interface is driven by a parabolic well of stiffness $m^2$ following some driving protocol $w(x,t)$. We restrict to monotonous driving $\dot w(x,t) \geq 0$ which leads to only forward motion $\dot{u}(x,t) \geq 0$
and to the so-called Middleton attractor. The lateral extension of the interface is noted $L >0$ and we assume periodic boundary conditions, although this will be unimportant as long as $L \gg 1/m$, the scale over which the interface motion is correlated. Our theory extends to other types of elasticity and more general microscopic disorder as in 
\cite{ThieryStaticCorrelations} but here we focus on this setting for the sake of simplicity. 

Our aim is to study the avalanches that occur in the so-called quasistatic limit. There are two main protocols that are largely equivalent. 
\begin{enumerate}
	\item
In the first protocol the interface driving is $w(x,t)=v t$ and we are interested in the stationary state \cite{middleton1992asymptotic} in the limit $v=0^+$ where the motion of the interface is intermittent. The avalanches are defined by the rapid motion $\dot u(x,t) \gg v$ that occur in between quiescence periods (of duration
of order $1/(L^d v)$).%{\red it should be $1/v$, why the factor $1/L^d$??}.
The avalanches can be indexed by the time at which they occur and their starting point ($x_i,t_i$) and we can ask about the correlations of these avalanches. Equivalently the process $u(x;w) = \lim_{v \to 0} u(x,t=w/v)$ exhibits jumps as a function of $w$ that are the avalanches. This process is called the quasi-static process and our goal is to study correlations between different given avalanches as a function of $W$, the distance along $u$ between two given avalanches, or equivalently ${\sf T} = W/v$, the time interval between the two avalanches (see Fig.~\ref{fig:ava1}).
Note that for $W>0$ many other avalanches have usually occured between the two avalanches under study.
%{\red I called it ${\sf T}$ everywhere not to confuse it with durations.}

\item

In the second protocol we prepare the system in the same stationary state, stop the driving at $w=0$ and wait for the interface to stop. It is thereby prepared with $\dot u(x,t=0)=0$ in the so-called Middleton attractor {\blue \cite{middleton1992asymptotic}}. Then we
apply a kick at $t=0$, $\dot w(x,t) = \delta w(x) \delta(t)$, either (i) local $\delta w(x)=\delta w \, \delta^d(x)$ or (ii) uniform $\delta w(x)=\delta w$.
This produces some interface motion, that we call an avalanche. Choosing uniform kicks of vanishing size $\delta w$ ensures that the interface stops at the
position $u(x; \delta w)$ previously defined in the first protocol in the limit $v \to 0^+$. To study the correlation between avalanches, we apply a series of such kicks, waiting each time for the interface to stop before applying the next kick. After $n \gg 1$ kicks the driving is now at $w = n \delta w$ and the position of the interface is $u(x;w)$.

\end{enumerate}

% We ask about the correlation of this sequence of avalanches. 

% It can be shown (see Appendix) that the sequence of avalanches in the first protocol, is statistically the same as the sequence obtained in the second protocol restricted to uniform kicks of vanishing sizes,
% with the correspondence $u(x;w)=u(x,t=w/v)$. This is what we call here the quasi-static process. 

For the sake of simplicity we use in this paper the language of the second protocol and more often consider uniform kicks (although the dependence in the positions of the avalanche starting points, the seeds, will be investigated using a local kick, see below). We derive results using the functional renormalization group and these will be accurate and universal in the limit of small $m$ (with still $L \gg 1/m$), in an expansion in $\epsilon = 4-d$ around the upper-critical dimension $d_{\rm uc} =4$ of the model. We thus restrict ourselves to $d<4$, although equivalent results for $d=4$ could be obtained.

\begin{figure}
\centering
\includegraphics[width=8cm]{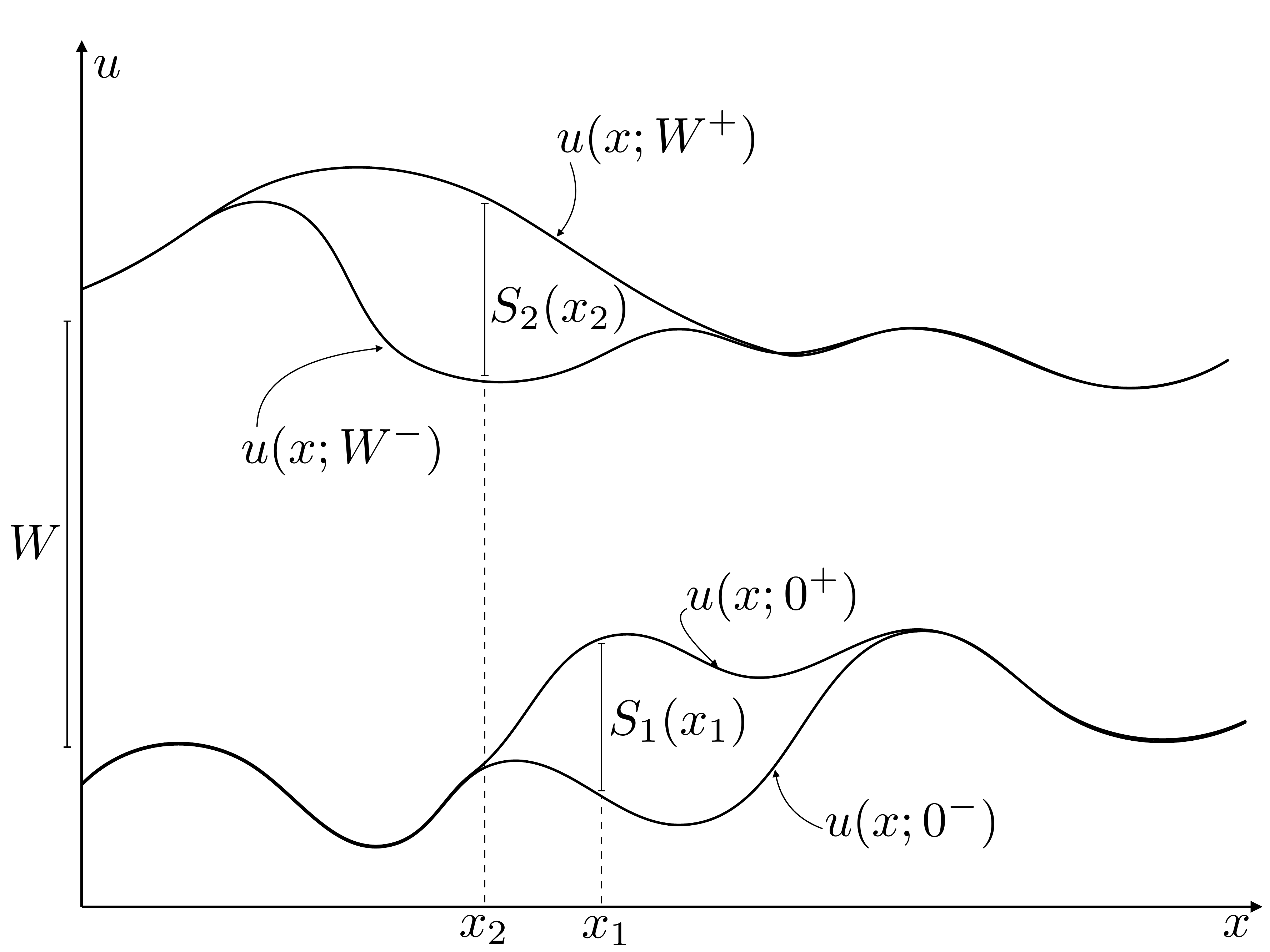}
\caption{The quasistatic process $u(x;w)$ for an elastic interface slowly driven in a disordered environment exhibits jumps called avalanches as a function of $w$ (the position of the driving well). Here an avalanche occured at $w=0$ and another one at $w=W$ (many others also occured in between since $u(x;0^+) \neq u(x,W^-)$). The goal of this article is to study the correlations between these two avalanches.
In particular the correlations between the total jumps (the areas of the two avalanches) or between
the local jumps $S_1(x_1)$ and $S_2(x_2)$ indicated on the figure.}
\label{fig:ava1}
\end{figure}

\subsection{Observables} \label{subsec:observables}

We focus on the correlations
between two avalanches, the first occuring after a uniform kick of size $\delta w_1$ at $w=0$ and the second occuring 
after a uniform kick of size $\delta w_2$ at $w=W$. If $W>0$ potentially many avalanches have also occured in between.
We start by considering the full velocity field
inside an avalanche. It is convenient to adopt the following representation.
We define two copies of the velocity field shifted in time $\dot u_1(x,\tau_1)$ and
$\dot u_2(x,\tau_2)$ where $\tau_1$ and $\tau_2$ denote the time since the
beginning of respectively the first and second avalanche (i.e. the first and the second kick, recalling
that, according to the second protocol described above there has been many kicks in between to move 
the interface from $w=0$ to $w=W$).

\subsubsection{Velocity fields during avalanches and their associated densities} \label{subsec:observables:velocity}

We consider a general observable of the form (i.e. the generating function of the velocity field)
\be
G[ \lambda] = \overline{ e^{\int d^dx dt \lambda(x,t) \dot{u}(x,t) }} \, .
\ee
Decomposing the source field $\lambda(x,t)$ as done above for the velocity, this can be rewritten 
%\be
%G[ \lambda] = G[ \lambda_1,\lambda_2 ] = \overline{ e^{\int dx [ \int_0^{+\infty} d\tau_1 \lambda_1(x,\tau_1) 
%\dot{u}_1(x,\tau_1) + \int_0^{+\infty} d\tau_2
%\lambda_2(x,\tau_2) \dot{u}_2(x,\tau_2)]}} 
%\ee
\be \label{defGen} 
G[ \lambda] = G[ \lambda_1,\lambda_2 ] = 
\overline{ e^{\lambda_1 \cdot \dot u_1 + \lambda_2 \cdot \dot u_2}} \, ,
\ee
where here and below we denote for each $i=1,2$
\be
\lambda_i \cdot \dot u_i = \int d^dx  \int_0^{+\infty} d\tau_i \lambda_i(x,\tau_i) 
\dot{u}_i(x,\tau_i) \, ,
\ee
where the source field $\lambda_1$ (respectively $\lambda_2$) is only non vanishing 
in the first (respectively the second) avalanche.\\

The generating function $G[\lambda]$ depends on the size of the kicks
$\delta w_i$ and can be expanded as follows
\be \label{expansion}
 G[ \lambda_1,\lambda_2 ] = 1 + \sum_{i=1,2} \delta w_i \int D \dot u_i 
\rho[\dot u_i] ( e^{\lambda_i \cdot \dot u_i}-1)   + \delta w_1 \delta w_2 
\int D[\dot u_1,\dot u_2]
\rho_W[\dot u_1,\dot u_2]
( e^{\lambda_1 \cdot \dot u_1}-1)(e^{\lambda_2 \cdot \dot u_2}-1)
  + O(\delta w_1^2, \delta w_2^2)  \, .
\ee
This is obtained by decomposing the process into events where either 
an avalanche occurs $\dot u>0$ or does not occur $\dot u=0$ 
\footnote{Here by an avalanche we mean a motion $\dot u =O(1)$. For $\delta w>0$
in the BFM there is always an avalanche, however most of them are of vanishing 
sizes and velocities as $\delta w \to 0^+$}.
The terms of order $\delta w_i$ account for the contribution to $G$ coming from events where an avalanche occured at $w=0$ or $w=W$, while the term of order $\delta w_1 \delta w_2$ accound for the contribution to $G$ where avalanches occured at both positions.
Here the factors $\rho[\dot u_i]$ denote the (equal by stationarity of the protocol considered here)
functional densities of the instantaneous velocity field taking the configuration $\dot u_i(x,\tau)$
during the corresponding avalanche (and $\int D \dot u_i$ denotes a functional integral). It is
normalized as $\int D \dot u_i 
\rho[\dot u_i] = \rho_0$ where $\rho_0 \sim L^d$  is the total density of avalanches per unit of driving $\delta w$. 
%The joint density $\rho_W[\dot u_1,\dot u_2]$ originates from
%events where an avalanche following the first and second kicks both occur. 
Similarly $\rho_W[\dot u_1,\dot u_2]$ is the joint density, i.e it is proportional to the number of events where two avalanches occured at $w=0$ and $w=W$ with velocity fields $\dot u_i(x,\tau)$. A more detailed discussion of the formula \eqref{expansion} is given in Appendix \ref{app:avalanches}.
If avalanches were independent, 
as is the case for the mean-field BFM (where they form a Levy jump process \cite{LeDoussalWiese2011b,ThieryLeDoussalWiese2015}) one would have
$\rho_W[\dot u_1,\dot u_2] = \rho[\dot u_1] \rho[\dot u_2]$. 
The present theory goes beyond the independent avalanche process, and allows to compute the
connected joint density
\be
\rho^c_W[\dot u_1,\dot u_2]  = \rho_W[\dot u_1,\dot u_2] - \rho[\dot u_1] \rho[\dot u_2] \, , 
\ee 
which vanishes in the mean field theory.

\subsubsection{Total and local avalanche size}   \label{subsec:observables:avalanchesize}

The theory presented in this paper allows to study any correlation between
the two velocity fields in the two avalanches. In this paper, to make calculations
and results explicit, we will first focus on the total and local size of the two avalanches
(the velocities being studied later in Section \ref{sec:dynamical}).
The local size of avalanche $i=1,2$ is defined as
the total displacement of the interface at a given point (see Fig.~\ref{fig:ava1})
\be \label{Sx} 
S_i(x) :=    \int_0^{+\infty} d\tau_i \dot{u}_i(x,\tau_i)  \, ,
\ee
and the total size is given by 
\be \label{S} 
S_i =  \int d^dx  S_i(x)  \, ,
\ee
that is the area spanned by the avalanche. 

The densities for these quantities can be obtained by considering the generating
function \eqref{defGen} for source fields chosen as $\lambda_i(x,\tau)= \lambda_i$,
for the total size, and as $\lambda_i(x,\tau)= \lambda_i \delta^d(x-x_i)$ 
for the local sizes $S(x_i)$. Expanding in powers of $\delta w_i$ gives, for the total size 
\bea \label{expansionS}
 && G[ \lambda_1,\lambda_2] = 1 +  \sum_{i=1,2} \delta w_i \int_0^{+\infty} dS_i 
\rho(S_i) ( e^{\lambda_i S_i}-1)   \\
&& + \delta w_1 \delta w_2 
\int_0^{+\infty} dS_1  \int_0^{+\infty} dS_2 
\rho_W(S_1,S_2)
( e^{\lambda_1 S_1}-1)(e^{\lambda_2 S_2}-1)
  + O(\delta w_1^2, \delta w_2^2)  \, ,
\eea
where $\rho(S)$ is the single avalanche total size density (per unit $w$)
normalized as $\int_0^{+\infty} dS S \rho(S)= 1$ and
$\rho_W(S_1,S_2)$ is the joint density of the total sizes $S_1,S_2$ in the two avalanches.
We also define the connected joint density as
\be
\rho^c_W(S_1, S_2)  = \rho_W(S_1,S_2) - \rho(S_1) \rho(S_2) \, ,
\ee 
which vanishes in mean field theory (the BFM) and, as we show below, 
is $O(\epsilon)$ where $\epsilon=d_c-d$, near the upper critical dimension 
$d_c=4$. Similarly, for the local size we have the same expansion
\eqref{expansionS} with $S_i \to S_i(x_i)$ and the corresponding
densities $\rho(S_i(x_i))$ and $\rho_W(S_1(x_1),S_2(x_2))$. In Sec.~\ref{subsec:seedcentered} we will also study the total size of avalanches conditioned on starting at a given point. The associated densities are defined in the same way.

\section{Dynamical field theory for velocity field correlations in two avalanches} 
\label{sec:analytical:fieldtheory}
\label{sec:analyticalresults}

We now present the dynamical field theory which allows us to calculate the
densities previously introduced to leading order in the $\epsilon=d_c-d$ expansion.
We also comment on the physical origin of the correlations.\\

\subsection{Field theory}

We now go back to consider the generating function $G[ \lambda_1,\lambda_2 ]$
in \eqref{defGen} for general sources. Our main result, justified in Appendix \ref{app:derivation}, is that, to
lowest order in an expansion around mean-field (i.e. independent avalanches), the generating function
which only measures the dynamics during the two avalanches separated by $W$ (see introduction)
can be written as
as the a functional average 
\be  \label{GenerFT1}
G%_{W,\delta w_1,\delta w_2}
[ \lambda_1,\lambda_2] = \int D[\tilde u_1,\dot u_1,\tilde u_2,\dot u_2]
e^{ \sum_{i=1,2} [ \lambda_i \cdot \dot u_i + 
m^2 \delta w_i \int d^d x \tilde{u}_i(x,\tau_i=0) ]
- S[\tilde u_1,\dot u_1,\tilde u_2,\dot u_2] }  \, ,
\ee
over the following dynamical action 
\be \label{ActionTotatle}
S[\tilde u_1,\dot u_1,\tilde u_2,\dot u_2]  = S_{BFM}[\tilde u_1,\dot u_1]
+ S_{BFM}[\tilde u_2,\dot u_2]
+ 
\Delta''(W)  \int d^d x \int_{\tau_1,\tau_2>0}
\tilde u_1(x,\tau_1) \tilde u_2(x,\tau_2) \dot u_1(x,\tau_1) \dot u_2(x,\tau_1)  \, .
\ee
Here $S_{\rm BFM}$ is the dynamical action associated to the BFM model
\be \label{ActionBFM}
S_{\rm BFM}[\dot{u} , \tilde{u}] = \int d^d x \left[ 
 \int_{\tau>0} \tilde{u}(x,\tau) (\eta \partial_\tau - \nabla^2 + m^2) \dot{u}(x,\tau)
-\sigma  \int_{\tau>0} \tilde{u}(x,\tau)^2 \dot{u}(x,\tau) \right]  \, ,
\ee
following a uniform kick $\delta w$ at time $\tau=0$. 
Here $\sigma=-\Delta'(0^+) > 0$ and here and below $\Delta(w)$ denotes the renormalized disorder
correlator defined from the two point corrrelation function of the position of the center of mass,
$u(w)=L^{-d} \int d^d x u(x,t=w/v)$, as
\be \label{defDelta}
\Delta(w-w')  = m^4 L^d \overline{ (u(w)-w)(u(w')-w') }^c  \, ,
\ee 
which is implicitly a function of $m$ and reproduces the bare disorder
correlator in the limit of large $m$, i.e. $\Delta_b(w)=\lim_{m \to +\infty} \Delta(w)$.
Here we focus on the universal $m \to 0$ limit, where
$\Delta(w)$ has been shown \cite{LeDoussalWieseChauve2002,RossoLeDoussalWiese2006a} to take the scaling form 
\be \label{scalingdelta} 
\Delta(w) =  A_d m^{\epsilon -  2 \zeta} \tilde \Delta(w m^\zeta) \quad , \quad A_{d=4} = 8 \pi^2 \quad , \quad 
\tilde \Delta^{* \prime \prime}(0^+)=\frac{1- \zeta_1}{3} \epsilon + O(\epsilon^2) 
\ee
with $A_d=2^{d-1}\pi^{d/2}/\Gamma(3-\frac{d}{2})$. Here $\tilde \Delta(w)$ converges as $m \to 0$ to the FRG fixed point $\tilde \Delta^*(w)$, which is uniformly of order $O(\epsilon)$
and solution of the FRG fixed
point equation, where $\zeta= \zeta_1 \epsilon + O(\epsilon^2)$ is the roughness exponent,
with $\zeta_1=1/3$ for interface depinning. Note that the depinning fixed point has one 
undetermined, non-universal constant $\kappa$, i.e. one can write $\tilde \Delta^*(w) = \kappa^2 \Delta^*(w/\kappa)$, however $\tilde \Delta^{* \prime \prime}(0^+)$ is fully universal, given in \eqref{scalingdelta}. The corresponding form for long-range elasticity can
be found in \cite{ThieryStaticCorrelations}.
To lowest order in $\epsilon$ the model is thus equivalent to two BFM models with a single
inclusion (at some fixed $\tau_1,\tau_2$) of the vertex $\Delta''(W)$ in \eqref{ActionTotatle} (see also \eqref{vertex0} below, the result being
later summed over all $\tau_1,\tau_2$). Diagrammatically it is represented
by two trees (the two BFM's) joined (i.e. correlated) by a single
vertex $\Delta''(W)$, as in Fig. 106 of \cite{ThieryStaticCorrelations}
(the same diagram holds for the dynamics, with time running upward). 

\medskip

The above theory allows to calculate exactly the correlations of the velocity field at
order $O(\delta w_1 \delta w_2)$ and to order $O(\epsilon)$
in the stationary setting defined in the introduction. Expanding 
Eq. \eqref{GenerFT1} in powers of $\delta w_1,\delta w_2$
and comparing with Eq. \eqref{expansion}, identifying the terms,
we obtain the Laplace transform 
of the single avalanche velocity field density in terms of the correlation of the response field as
\be \label{LTVelocity1}
 \int D \dot u_1
\rho[\dot u_1] [ e^{\lambda_1 \cdot \dot u_1 }-1]  
= m^2 \int D[\tilde u_1,\dot u_1] \int dx \tilde u_1(x,0) 
e^{ \lambda_1 \cdot \dot u_1  - S_{\rm BFM}[\tilde u_1,\dot u_1]} \, ,
\ee
as well as the joint density of velocity fields in the two avalanches, as
\bea \label{rhoWform} 
&& \int D[ \dot u_1,\dot u_2]
\rho_W[\dot u_1,\dot u_2] 
( e^{\lambda_1 \cdot \dot u_1}-1)(e^{\lambda_2 \cdot \dot u_2}-1)
 \\
&& 
= m^4  \int D[\tilde u_1,\dot u_1,\tilde u_2,\dot u_2]
 \int dx_1 \tilde u_1(x_1,0) \int dx_2 \tilde u_2(x_2,0) 
e^{  \sum_{i=1,2} \lambda_i \cdot \dot u_i  - S[\tilde u_1,\dot u_1,\tilde u_2,\dot u_2]}   \, . \nn 
\eea

We now make these results more explicit by focusing on
simpler observables, namely the joint densities and correlations of the total sizes $S_1,S_2$, as well
as the local sizes $S_1(x), S_2(y)$ of the two avalanches.

\medskip 

{\it Remark 1:}  To be more precise, we stress that the above theory is devised to calculate the 
$O(\epsilon)$ result for the connected correlations between two avalanches.
For the velocity field statistics inside a single avalanche it only leads to the
mean field result, i.e. $O(\epsilon^0)$. To obtain the $O(\epsilon)$ correction
to the latter one needs to add other terms (involving $\Delta''(0)$) in the action
as was done in \cite{LeDoussalWiese2012a} (see also Appendix \ref{app:derivation}). This is not our purpose here, and one should remember that only the results that we obtain for the {\it connected} correlations (the sole purpose of this paper) between avalanches are exact up to order $O(\epsilon)$.

\medskip

{\it Remark 2:} The above setup and results can be easily generalized to study correlations between avalanches that are conditioned on starting at any given position (called the `seed') along the interface. Replacing $\int dx_1 \tilde u_1(x_1,0) \int dx_2 \tilde u_2(x_2,0) \to \tilde u_1(x_1,0)  \tilde u_2(x_2,0) $ in \eqref{rhoWform} 
for given $x_1$ and $x_2$ fixed, indeed selects the avalanches that have started at $x=x_1$ at $w=0$ and at $x=x_2$ at $w=W$. The easiest way to see this is to slightly modify our protocol by triggering the avalanches at $w=0$ and $w=W$ by local kicks $\delta w_i(x) =  \delta^d(x-x_i) \delta w_i$ of vanishing size $\delta w_i$. Such kicks can indeed only trigger avalanches with seeds $x_1$ and $x_2$. One then easily generalizes 
\eqref{expansionS} to this case, see also Appendix \ref{app:avalanches}. This is used in Sec.~\ref{subsec:seedcentered}. We refer the reader to \cite{ThieryShape,ThieryPhD} for more details about this seed centering procedure in the field theory.

\medskip

{\it Remark 3:} One can study truly successive avalanches by simply considering the
limit $W=0^+$. The present theory applies as long as the second avalanche starts
well after the first one is finished. It would be
interesting to study overlapping avalanches, but this goes beyond this paper
(see discussion in Appendix \ref{app:avalanches}).\\

\subsection{Origin of correlations}

Before we go further here let us comment here on the physical origin of the correlations. The correlations originate from the fact that the time derivatives of
the pinning forces $\partial_{\tau_i} F(u_i(x,\tau_i))$, which enters the equation
of motion for the velocity, are correlated, their covariance
being
\bea \label{vertex0} 
\overline{\partial_{\tau_1} F(u_1(x,\tau_1),x) \partial_{\tau_2} F(u_2(x',\tau_2),x') }
&= & \partial_{\tau_1} \partial_{\tau_2} \Delta(u_1(x,\tau_1) - u_2(x,\tau_2) + W) \delta^d(x-x') \\
& \simeq & - \dot u_1(x,\tau_1) \dot u_2(x,\tau_2) \Delta''(W) \delta^d(x-x') \nonumber 
\eea 
This is because two displacement fields in the two avalanches see the same 
static random pinning force landscape. This landscape is correlated in
the direction of the motion, and the correlation, at fixed value of $u_1(x,\tau_1) - u_2(x,\tau_2)$
extend to arbitrary time difference (we recall that $\tau_i$ are counted from the beginning times
${\sf T}_i$ of each avalanche, e.g. hence $u_i(x,\tau_i)=u(x,t={\sf T}_i+\tau_i)$,
where ${\sf T}_2-{\sf T}_1 \sim W/v$ is a very large time). \\

One could object, however, that the SR correlator
of the bare (i.e. microscopic) model, $\Delta_b(w)$, has correlations only on short distance
$w \sim r_f$, leading to correlations of the displacements only within the
(small and fixed) Larkin volume. However, the {\it renormalized} correlator, $\Delta(w)$, which
includes the effect of the interplay of elasticity and disorder at all scales
is correlated on the much larger distance $w \sim m^{-\zeta}$.
It is this renormalized correlator that must be used here. It is
possible to prove this fact (and also justify that it is $\sigma=-\Delta'(0^+)$ which
must be used in the BFM) by consideration of the effective action
of the theory. Some steps in that direction are provided in the Appendix~\ref{app:derivation}. Physically, correlations live on the scale $w \sim m^{-\zeta}$ because this is the scale of the displacement of the interface during avalanches. In the dynamics we always find that the correlations are negative. It can be seen
already from the negative correlation of the forces in \eqref{vertex0} (since $\Delta''(W)>0$ near the
depinning fixed point and the velocities are positive). One finds that if one avalanche occured and the interface moved on a distance $m^{-\zeta}$, the driving has to `catch up' with this scale $w \sim m^{-\zeta}$ in order for the interface to forget that this avalanche has already occured.\\

\section{Analytical results: Correlations of the avalanche sizes, total and local}
\label{sec:size}

\subsection{Joint density of total sizes}  \label{subsec:analytical:totalsize}

We first consider the total sizes of the avalanches defined in \eqref{S}.

\subsubsection{Expressions for Laplace tranform}

Following the same steps as in the previous section, expanding 
\eqref{GenerFT1} in powers of $\delta w_1,\delta w_2$ in the special
case of constant sources $\lambda_i(x,\tau)= \lambda_i$
and comparing with \eqref{expansionS} we obtain 
the Laplace transform of $\rho(S)$ as 
\be \label{rho1} 
 \int dS_1 
\rho(S_1) [ e^{ \lambda_1 S_1 }-1]  
 = m^2  \int D\tilde u_1 D\dot u_1 \int dx \tilde u_1(x,0) 
e^{ \lambda_1 \int dx  \int_0^{+\infty} d\tau_1  
\dot{u}_1(x,\tau_1)  - S_{BFM}[\tilde u_1,\dot u_1]} \, ,
\ee
a formula similar to \eqref{LTVelocity1}, and the Laplace transform of $\rho_W(S_1,S_2)$ as 
\bea \label{rho2}
&& \int dS_1 dS_2
\rho_W(S_1,S_2) [ e^{ \lambda_1 S_1 }-1]  [ e^{ \lambda_2 S_2 }-1]    \\
&& 
=  m^4 \int D\tilde u_1 D\dot u_1 \int D\tilde u_2 D\dot u_2 
 \int dx_1 \tilde u_1(x_1,0) \int dx_2 \tilde u_2(x_2,0) 
e^{\sum_{i=1,2} \lambda_i \int d^d x \int_0^{+\infty} d\tau_i \dot u_i(x,\tau_i)   - S[\tilde u_1,\dot u_1,\tilde u_2,\dot u_2]} \nn \, ,
\eea
a formula similar to \eqref{rhoWform}.
We now compute explicity the r.h.s. of these equations. 

\subsubsection{Review of the calculation of $\rho(S)$}

The r.h.s of \eqref{rho1} can be obtained from a standard 
calculation within the BFM. The main observation is that the field
$\dot u_1$ in the exponential on the r.h.s. of \eqref{rho1}  appears only linearly \cite{LeDoussalWiese2012a}. Hence integrating over it leads
to a delta function which constrains $\tilde u(x,t)$ to be a solution
of the so-called instanton equation equation \cite{LeDoussalWiese2012a}, in the present case given by
\bea
\tilde u(x,t) = \tilde u \quad , \quad - m^2 \tilde u + \sigma \tilde u^2
= - \lambda \, ,
\eea 
which is
\bea
\tilde u = \frac{1}{m^2 S_m} Z(\lambda S_m) \quad , \quad Z(\lambda)= 
 \frac{1}{2} ( 1- \sqrt{1 - 4 \lambda} ) \, ,
\eea 
where $S_m=\frac{\sigma}{m^4}$ is the typical scale of the largest avalanches
in the BFM. This leads to 
\be \label{rho11} 
 \int dS_1 
\rho(S_1) [ e^{ \lambda_1 S_1 }-1]  = \frac{1}{S_m} L^d Z(S_m \lambda) \, , 
\ee
which leads to the total size density
\be \label{eq:defhatrho}
\rho(S)= \frac{L^d}{S_m^2} \hat \rho(S/S_m) \quad , \quad
\hat \rho(s) = \frac{1}{2 \sqrt{\pi} s^{3/2}} e^{-s/4}  \, ,
\ee
which is the classic result for the BFM (exact in our setting at order $O(\epsilon^0)$).
Note that $\int_0^{+\infty} ds s \hat \rho(s) =1$ and $\int_0^{+\infty} ds s^2 \hat \rho(s) =2$.

\subsubsection{Calculation of $\rho_W(S_1,S_2)$} \label{sec:calculation1} 

The r.h.s of \eqref{rho2} can be obtained from a modification of
the previous calculation. The difficulty is the term 
proportional to $ \Delta''(W) \dot u_1 \dot u_2$ in the action $S$ in Eq. \eqref{ActionTotatle}.
It can be decoupled by the following calculational trick. We introduce formal centered Gaussian noise fields 
$\xi_i(x,t)$ as 
\bea
e^{- \Delta''(W)  \int d^d x \int_{\tau_1,\tau_2>0}
\tilde u_1(x,\tau_1) \tilde u_2(x,\tau_2) \dot u_1(x,\tau_1) \dot u_2(x,\tau_1) }
= \langle 
e^{\sum_{i=1,2} \int d^dx \int_0^{+\infty} d\tau_i \xi_i(x)  \tilde u_i(x,\tau_i)  \dot u_i(x,\tau_i) }
\rangle_{\xi}  \, ,
\eea
where by definition
\be \label{x1x2} 
 \langle \xi_i(x) \xi_j(x') \rangle_{\xi} = - (1-\delta_{ij}) \Delta''(W) \delta^d(x-x') \, .
\ee
For a given noise $\xi_i(x)$, the velocity fields now appear linearly and one can integrated over them  (as in the BFM). This, for each
realization of the noise $\xi$, constrains the response fields $\tilde u_i(x,t)$ to obey 
two decoupled instanton equations, whose solutions are time independent but space 
inhomogeneous, $\tilde u_i(x,t)=\tilde u_i(x)$,
where $\tilde u_i(x)$ for $i=1,2$ are solution of
\bea \label{inst2} 
 ( \nabla_x^2 - m^2) \tilde u_i(x) + \sigma \tilde u_i(x)^2 = - \xi_i(x) \tilde u_i(x) 
 - \lambda_i  \, .
\eea 
The solutions of these equations are coupled because the noises $\xi_1$ and $\xi_2$
are not independent, and correlated as in \eqref{x1x2}. From these
solutions, and from \eqref{rho2}, one obtains the Laplace transform of the
joint density as
\be \label{rho22}
 \int dS_1 dS_2
\rho_W(S_1,S_2) [ e^{ \lambda_1 S_1 }-1]  [ e^{ \lambda_2 S_2 }-1]    
=  m^4 \int d^d x_1\int d^d x_2 \langle \tilde u_1(x_1)  \tilde u_2(x_2) \rangle_\xi \, .
\ee
Being interested in computing $\rho_W(S_1,S_2)$ in first order in $\Delta''(W)$
(which is itself $O(\epsilon)$)
implies that we only need to solve perturbatively \eqref{inst2} to first order in
$\xi$. To this order the solutions can be written as
\bea
\tilde u_i(x)  = \tilde u^0_i +  \tilde u^1_i(x)  \quad , \quad \tilde u^0_i = \frac{1}{m^2 S_m} Z(S_m \lambda_i) \, ,
\eea 
and in Fourier space
\bea \label{u1} 
\tilde u^1_i(q) = \frac{1}{m^2 S_m} \, \frac{Z(S_m \lambda_i)}{q^2 
+ m^2 - 2 m^2 Z(S_m \lambda_i)} \xi_i(q)  \, .
\eea 
Leading to our main result for the connected joint density
\be \label{rho22}
 \int dS_1 dS_2
\rho^c_W(S_1,S_2) [ e^{ \lambda_1 S_1 }-1]  [ e^{ \lambda_2 S_2 }-1]    
=  - \Delta''(W) 
\frac{L^d}{m^4 S_m^2} \, \frac{Z(S_m \lambda_1)}{1- 2 Z(S_m \lambda_1)}
\, \frac{Z(S_m \lambda_2)}{1 - 2  Z(S_m \lambda_2)} \, ,
%\frac{Z(\lambda_1)}{1-2Z(\lambda_1)} \frac{Z(\lambda_2)}{1-2Z(\lambda_2)}
\ee
where the part proportional to $\tilde u_1^0 \tilde u_2^0$ cancels in the
connected density.

The formula \eqref{rho22} if formally identical to the result Eq. (71)
in \cite{ThieryStaticCorrelations} for the statics. This shows that the present theory reproduces the results of \eqref{rho22} in a much simpler fashion. However, one must stress that the 
renormalized correlator $\Delta(W)$ is different in the dynamics
from its value e.g. for random bond statics, leading to a numerically different result.

By expanding \eqref{rho22} in powers of $\lambda_i$ one obtains
the integer moments over $\rho^c_W$ which we denote
$\langle \dots \rangle_W^c$. Similarly the averages over the single avalanche
density $\rho$ are denoted as $\langle \cdots \rangle$ \footnote{Since below
we only consider moment ratio the global normalization drops out and we can
define $\langle \cdots \rangle_W^c = \int dS_1 dS_2 \cdots \rho^c_W(S_1,S_2)$
and $\langle \cdots \rangle= \int dS  \cdots \rho(S)$.}. We give here two explicit formula, which are
tested in the numerics in Sec.~\ref{sec:numerics} below. First one finds
\bea \label{eq:resS1S2}
\frac{ \langle S_1 S_2 \rangle_W^c}{ \langle S \rangle^2}=- \frac{\Delta''(W)}{m^4 L^d} \, ,
\eea 
which is in fact an exact result, and can be seen to follow from
the definition \eqref{defDelta} of the renormalized disorder correlator.
The proof is identical to the statics case, to which we refer
(Eq. (8) and Section III.F. and IV. E. in \cite{ThieryStaticCorrelations}). 
Another result is 
\bea \label{eq:resS1S22}
\frac{ \langle S_1^2 S_2 \rangle_W^c}{ \langle S^2 \rangle \langle S \rangle}
= \frac{ \langle S_1 S_2^2 \rangle_W^c}{ \langle S^2 \rangle \langle S \rangle}
=  - 3 \frac{\Delta''(W)}{m^4 L^d} \, ,
\eea 
which holds only to order $O(\epsilon)$. 
Note that to the order that we calculate here 
$\langle S_1^2 S_2 \rangle_W^c= \langle S_1 S_2^2 \rangle_W^c$
and more generally, at this order the correlation between the two avalanches
are symmetric (which is likely not to hold to higher orders in the $\epsilon$ expansion
contrary to the statics). 

As in \cite{ThieryStaticCorrelations}, \eqref{rho22} can be inverse Laplace transformed, and leads to the complete $O(\epsilon)$ result for the connected density
\bea \label{rhocW} 
\rho_W^c(S_1,S_2) = - \frac{\Delta''(W)}{L^d m^4} \frac{S_1 S_2}{4 S_m^2} \rho(S_1) \rho(S_2) \, + O(\epsilon^2) \, ,
\eea
which again, is formally identical to (10) in  \cite{ThieryStaticCorrelations}. We recall the definition
\be \label{defSm}
S_m= \frac{|\Delta'(0^+)|}{m^4} = \frac{\langle S^2 \rangle}{2 \langle S \rangle}
\ee
valid beyond mean-field. We can rewrite the connected density in the form
\bea \label{scalform}
\rho_W^c(S_1,S_2) = \frac{1}{(L m)^d} \frac{L^{2d}}{S_m^4} {\cal F}_d(\frac{W}{W_m},\frac{S_1}{S_m},\frac{S_2}{S_m})
\eea
where ${\cal F}_d$ is a universal function, with
\bea \label{scaluniform}
{\cal F}_d(w,s_1,s_2) \simeq - \frac{A_d \tilde \Delta^{* \prime \prime}(w)}{16 \pi \sqrt{s_1 s_2}} e^{-(s_1+s_2)/4} + O(\epsilon^2) \, ,
\eea
correcting the sign misprint in (13), (89) and (90) of \cite{ThieryStaticCorrelations}, 
where the prefactor is given by \eqref{scalingdelta}, and 
fully universal at $w=0$, i.e. $\Delta^{* \prime \prime}(0^+)=\frac{2}{9} \epsilon$ and $A_4=8 \pi^2$.
The scale $W_m \simeq \kappa m^{-\zeta}$ contains one non-universal amplitude,
related however to $S_m = - \Delta'(0^+)/m^4 = - A_d \tilde \Delta'(0^+) m^{-d-\zeta}
\simeq A_d \kappa \epsilon \sqrt{1-2 \zeta_1} m^{-d-\zeta}$ (see formula (90) and below in \cite{ThieryStaticCorrelations})
which can be independently measured from \eqref{defSm}, allowing to determine $\kappa$.\\

Finally, note that for any real $p_1,p_2,q_1,q_2>-1/2$ with $p_1+p_2=q_1+q_2$ we predict the 
following dimensionless ratio
\bea
\frac{ \langle S_1^{p_1} S_2^{p_2} \rangle_W^c}{\langle S_1^{q_1} S_2^{q_2} \rangle_W^c}
= \frac{\Gamma(p_1+\frac{1}{2}) \Gamma(p_2+\frac{1}{2})}{\Gamma(q_1+\frac{1}{2})\Gamma(q_2+\frac{1}{2})}    \,
\eea

\subsection{Joint density of local sizes} \label{subsec:analytical:localsize}

\subsubsection{Dimensionless units}

In order to lighten notations and calculations, from now on we switch to dimensionless units. We introduce the characteristic scales of avalanches. The lateral extension $1/m$, total size $S_m = \sigma/m^4$, duration $\tau_m = \eta/m^2$, velocity $v_m = m^d  S_m/\tau_m$. We rescale space and time as $x \to x/m$, $t \to \tau_m t$. The fields are rescaled as $u \to m^d S_m u$, $\dot u \to v_m \dot u$, $\tilde u \to  \frac{1}{m^2 S_m} \tilde{u}$. Avalanche local and total sizes are rescaled accordingly as $S(x) \to m^d S_m S(x)$, $S \to S_m S$. We also use that the renormalized disorder correlator $\Delta$ takes a scaling form $\Delta(W) =  m^{\epsilon -  2 \zeta} \hat{\Delta}(m^\zeta W)$ with $\zeta$ the roughness exponent of the interface. We rescale the distance between avalanches as $W \to m^{-\zeta} W$. This is equivalent to setting $\eta = m = \sigma =1$ in the above theory with also the replacement $\Delta''(W) \to \hat{\Delta}''(W)$. We will reintroduce the full dimensions explicitly for some results,
which can be done easily using Appendix \ref{app:restore}.

\subsubsection{Expressions for Laplace tranforms}

To study correlations between avalanche local sizes $S_1(x_1)$ and $S_2(x_2)$, see 
Fig. \ref{fig:ava1},
we now do as in Sec.~\ref{subsec:analytical:totalsize} but using source fields $\lambda_1(x,\tau_1)  = \lambda_1 \delta^d(x-x_1)$ and $\lambda_2(x,\tau_2) = \lambda_2 \delta^d(x-x_2)$. Expanding 
\eqref{GenerFT1} in powers of $\delta w_1,\delta w_2$ in the special
case of these sources we obtain 
the Laplace transform of $\rho(S_1(x_1))$ as (a formula similar to \eqref{LTVelocity1})
\be  \label{LTrhoSx}
 \int dS_1(x_1)
\rho(S_1(x_1)) [ e^{ \lambda_1 S_1(x_1) }-1]  
 = \int D\tilde u_1 D\dot u_1 \int dx \tilde u_1(x,0) 
e^{ \lambda_1   \int_0^{+\infty} d\tau_1 
\dot{u}_1(x_1,\tau_1)  - S_{BFM}[\tilde u_1,\dot u_1]} \, ,
\ee
and of $\rho_W(S_1(x_1),S_2(x_2))$ as (a formula similar to \eqref{rhoWform})
\bea \label{LTrhoS1x1S2x2}
&& \int dS_1(x_1) dS_2(x_2)
\rho_W(S_1(x_1),S_2(x_2)) [ e^{ \lambda_1 S_1(x_2) }-1]  [ e^{ \lambda_2 S_2(x_2) }-1]    \\
&& 
=  \int D\tilde u_1 D\dot u_1 \int D\tilde u_2 D\dot u_2 
 \int dy_1 \tilde u_1(y_1,0) \int dy_2 \tilde u_2(y_2,0) 
e^{\sum_{i=1,2} \lambda_i  \int_0^{+\infty} d\tau_i \dot u_i(x_i,\tau_i)   - S[\tilde u_1,\dot u_1,\tilde u_2,\dot u_2]}  \, .\nn
\eea
We now compute explicity the r.h.s. of these equations following the same procedure as in the previous section.

\subsubsection{Review of the calculation of $\rho(S(x))$}

As in Sec.~\ref{subsec:analytical:totalsize}, it can be seen that $\dot u_i$ only appears linearly in the exponential in the rhs of \eqref{LTrhoSx}. Integrating over it creates a Dirac delta functional and constrains the response field $\tilde{u}_i$ as $\tilde{u}_i(x,\tau_i) = \tilde{u}_i(x_i)$ with $\tilde{u}_i(x_i)$ the solution of the following space inhomogeneous instanton equation:
\bea \label{inst0}
(\nabla_x^2 - 1) \tilde u_i(x) +  (\tilde u_i(x))^2 = - \lambda_i \delta^d(x-x_i) \, .
\eea 
The Laplace transform of $\rho(S(x_i))$ is then computed as, using \eqref{LTrhoSx} 
\be \label{rho1x} 
 \int dS(x_i) 
\rho(S(x_i)) [ e^{ \lambda_i S(x_i) }-1]  = \int d^dx \, \tilde u_i(x)  \, .
\ee
As discussed in \cite{LeDoussalWiese2008c,LeDoussalWiese2012a,Delorme} the solution $ \tilde u_i(x) $ can be exactly obtained in $d=1$ as

\be  \label{Yscaling2}
\tilde u_i(x) = \tilde u_i^0(x) := \frac{6(1 - z_i^2) e^{- |x-x_i| }  }{\left( 1 + z_i + (1-z_i) e^{- |x-x_i| }  \right)^2}\ ,
\ee 
where $z_i(\lambda_i)$ is one of the solutions of 
\be  \label{eqzi}
\lambda_i = 3 z_i (1 -z_i^2)  \ . 
\ee 
The right solution satify the following properties: it is defined for $ \lambda_i \in ] - \infty  ,  2/\sqrt{3} [$, decreases from $z_i(-\infty) = \infty$ to $z_c = z_i (2/\sqrt{3} )= 1/\sqrt{3}$ and approaches $1$ as $\lambda_i$ approaches $0$.
It is possible to perform the Laplace inversion leading to \cite{LeDoussalWiese2008c,LeDoussalWiese2012a,Delorme}
\be
\rho(S(x_i)=S_0) = \frac{2}{\pi S_0} K_{1/3}(\frac{2 S_0}{\sqrt{3}})  \, .
\ee 
%We refer to *** for more details on the resulting distribution $\rho(S(x_i))$.

\subsubsection{Calculation of $\rho_W(S_1(x_1),S_2(x_2))$}

To compute $\rho_W(S_1(x_1),S_2(x_2))$ we follow the same steps as in Sec. \ref{sec:calculation1} and linearize in the $\dot{u}_i$ fields the argument of the exponential in the r.h.s. of \eqref{LTrhoS1x1S2x2} by reintroducing the formal Gaussian fields $\xi_i(x)$. This leads to the expression 
\bea \label{rho23}
&& \int dS_1(x_1) dS_2(x_2)
\rho_W(S_1(x_1),S_2(x_2)) [ e^{ \lambda_1 S_1(x_1) }-1]  
[ e^{ \lambda_2 S_2(x_2) }-1]   = \langle \int_{y_1,y_2} d^dy_1 d^dy_2 \tilde{u}_1(y_1) \tilde{u}_2(y_2) \rangle_\xi  \, ,
\eea
in terms of the solution of the following space inhomogeneous instanton equation 
\bea \label{inst22} 
 ( \nabla_x^2 - 1) \tilde u_i(x) + \tilde u_i(x)^2 = -\xi_i(x) \tilde u_i(x)
 - \lambda_i \delta^d(x-x_i)  \, .
\eea 
Again to obtain the result at order $O(\epsilon)$ it is sufficient to solve perturbatively \eqref{inst22} to first order in
$\xi$. This leads to
\bea
\tilde u_i(x)  = \tilde u_i^0(x) + \int d^dy \, G_i(x,y) \xi_i(y) \tilde u_i^0(y)   + O(\xi^2) \, ,
\eea 
where $\tilde{u}_i^0(x)$ is the solution of \eqref{inst0}, which in $d=1$ is given in \eqref{Yscaling2}. We have introduced the propagators $G_i(x,y)$ satisfying the following equations 
%{\red I changed the definition with a minus sign, and corrected all other equations}
\bea  \label{defpropag}
 ( \nabla_x^2 - 1 + 2 \tilde u_i^0(x) ) G_i(x,y) = - \delta^d(x-y)  \, .
 \eea 

 Inserting this formal solution in \eqref{rho23} and using the noise correlations \eqref{x1x2}, we finally obtain the Laplace transform of the connected density for the local size as 

\bea \label{rho24}
&& \int dS_1(x_1) dS_2(x_2)
\rho^c_W(S_1(x_1),S_2(x_2)) [ e^{ \lambda_1 S_1(x_1) }-1]  
[ e^{ \lambda_2 S_2(x_2) }-1]    
\\
&& = - \hat{\Delta}''(W)
\int d^d z_1 \int d^d z_2 
\int d^dy G_1(z_1,y) G_2(z_2,y) \tilde u^0_1(y) \tilde u^0_2(y) \, .
\eea
Again, it can be seen that this result reproduce the equivalent results obtain for shocks in the static. This is most easily seen by comparing this expression with the expression (D16) in Appendix D of \cite{ThieryStaticCorrelations}.

From these expressions one easily obtain by expanding in $\lambda_i$ a few integer moments of the connected distribution (see \cite{ThieryStaticCorrelations}). 
Here we only give the explicit result for the first moment in $d=1$ that will be compared with numerical simulations in Sec.~\ref{sec:numerics}.  One easily obtains from \eqref{eqzi} that $z_i(\lambda_i) = 1 - \frac{1}{6} \lambda_i + O(\lambda_i^2)$. Then from \eqref{Yscaling2} one obtains $\tilde{u}_i(x) = \frac{\lambda_i}{2} e^{-|x-x_i|} + O(\lambda_i^2)$. From \eqref{defpropag} one obtains $G_i(x,y) = \frac{1}{2} e^{-|x-y|} + O(\lambda)$ and thus we obtain from \eqref{rho24} that
\bea
\langle S_1(x_1) S_2(x_2) \rangle^c_W && =_{d=1} - \hat{\Delta}''(W) \int dz_1 dz_1 dy  \frac{1}{16} e^{- |z_1 -y| - |z_2-y| - |y-x_1| - |y-x_2|} =   - \frac{\hat{\Delta}''(W)}{4}  \int dy  e^{ - |y-x_1| - |y-x_2|} \nn \\
&& = - \frac{\hat{\Delta}''(W)}{4} ( 1+ |x_2-x_1|)  e^{-|x_2-x_1|} \, .
\eea 
Reintroducing the units and normalizing, one gets
\bea \label{eq:resS1x1S2x2}
 \frac{\langle S_1(x_1) S_2(x_2) \rangle^c_W}{\langle S(x) \rangle^2} =_{d=1}  - m^{-3} \frac{\Delta''(W)}{4} ( 1+ m|x_2-x_1|)  e^{-m|x_2-x_1|} \, ,
\eea
a result that reproduces the result (112) of \cite{ThieryStaticCorrelations}.

\subsection{Joint density of total sizes for given positions of the seeds} \label{subsec:seedcentered}

\begin{figure}[h]
\centering
\includegraphics[width=8cm]{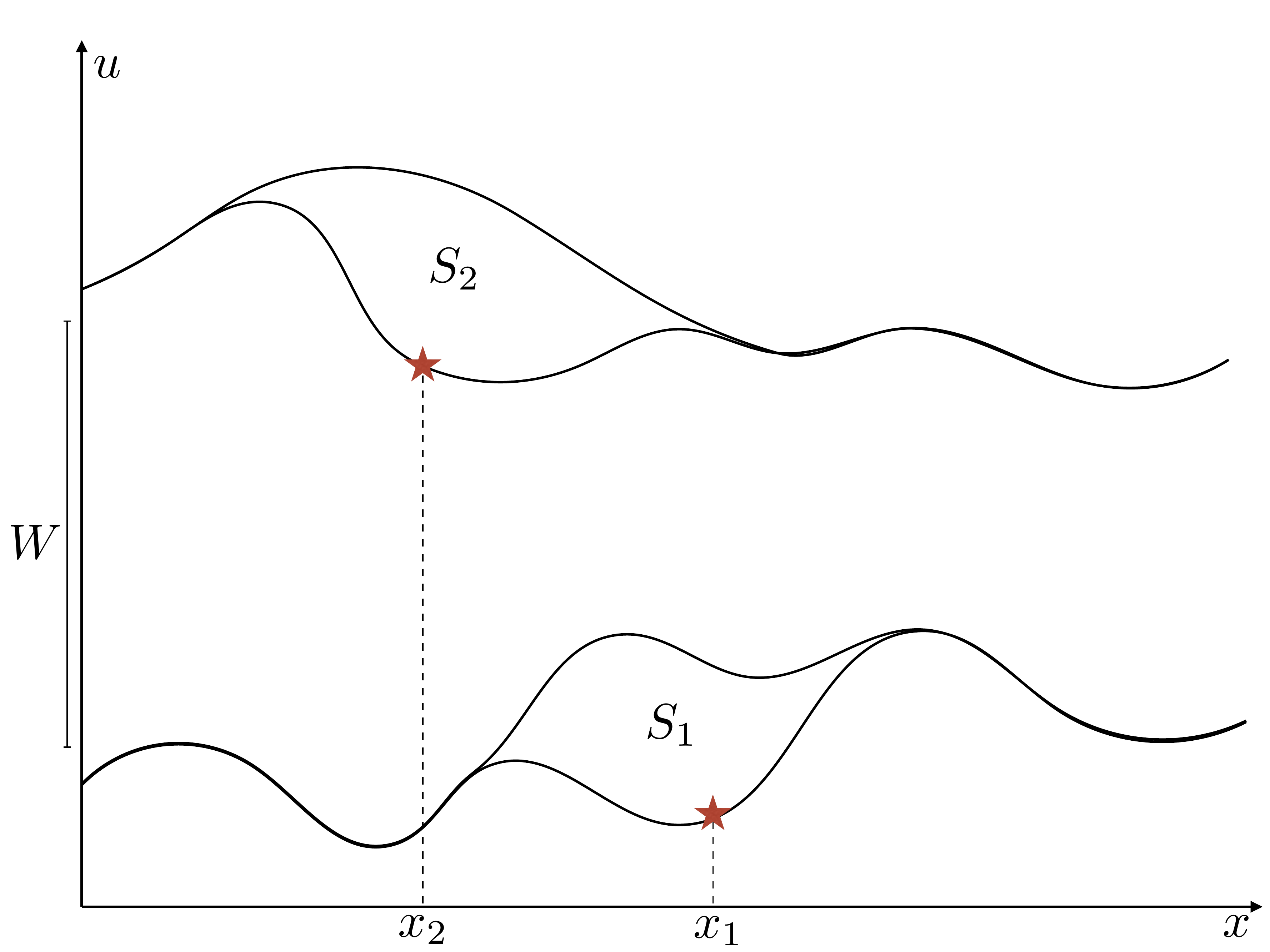}
\caption{In Section \ref{subsec:seedcentered} we calculate the correlations between the total sizes $S_1$ and $S_2$ of two avalanches separated by a distance $W$ along $u$, with seeds (i.e. starting points) at $x_1$ and $x_2$.}
\label{fig:ava2}
\end{figure}

\subsubsection{Seed centering and Laplace tranform}

The explicit results we have obtained up to know are formally equivalent to the results obtained for shocks in the statics \cite{ThieryStaticCorrelations}, rederived in the dynamics in a much simpler fashion using our general result \eqref{GenerFT1}-\eqref{ActionTotatle}-\eqref{ActionBFM}. We now obtained some genuinely new results by looking at the correlations between the total size $S_1$ and $S_2$ of avalanches occuring at a distance $W$ and having seeds $x_1$ and $x_2$ (see Fig.~\ref{fig:ava2}). By `seed' we mean the first point of the interface that moves when the avalanche starts. This concept is much more natural in the
dynamics than in the statics. We now introduce $\rho^x(S)$ and $\rho^{x_1,x_2}(S_1,S_2)$ the total size density for avalanche that starts at $x$ and total size density for avalanches that starts at $x_1$ at $w=0$ and at $x_2$ at $w=W$. By translational invariance along the internal direction we have 

\be
\rho^{x}(S) = \frac{1}{L^d} \rho(S) \quad , \quad \rho^{x_1,x_2}_W(S_1,S_2) \equiv  \rho^{x_2-x_1}_W(S_1,S_2)  \, ,
\ee
with the normalization 
\be
\int d^d x \rho^{x}_W(S_1,S_2) = \frac{1}{L^d} \rho_W(S_1,S_2)  \, .
\ee

Our theory predicts that the Laplace transform of these densities are obtained as in \eqref{rho1}-\eqref{rho2} but with the response fields in front of the exponential not integrated over but restricted at $x_1$ or $x_2$ (see remark in Sec.~\ref{subsec:analytical:fieldtheory} for a justification). More precisely we get
\be \label{rho1new} 
 \int dS_1 
\rho^{x_1}(S_1) [ e^{ \lambda_1 S_1 }-1]  
 = \int D\tilde u_1 D\dot u_1 \tilde u_1(x_1,0) 
e^{ \lambda_1 \int dx  \int_0^{+\infty} d\tau_1 
\dot{u}_1(x,\tau_1)  - S_{BFM}[\tilde u_1,\dot u_1]} \, ,
\ee
and 
\bea \label{rho2new}
&& \int dS_1 dS_2
\rho_W^{x_1,x_2}(S_1,S_2) [ e^{ \lambda_1 S_1 }-1]  [ e^{ \lambda_2 S_2 }-1]    \\
&& 
=   \int D\tilde u_1 D\dot u_1 \int D\tilde u_2 D\dot u_2 
 \tilde u_1(x_1,0) \tilde u_2(x_2,0) 
e^{\sum_{i=1,2} \lambda_i \int d^d x \int_0^{+\infty} d\tau_i \dot u_i(x,\tau_i)   - S[\tilde u_1,\dot u_1,\tilde u_2,\dot u_2]} \nn \, .
\eea

\subsubsection{Calculation of the densities}
\label{sec:calc} 

Proceeding as in Sec.~\ref{subsec:analytical:totalsize}, it is easily seen that the seed centering procedure does not affect the instanton equations satisfied by the response fields and, as mentioned above, the only modification that is necessary is to keep track of the spatial dependence of the solutions of the instanton equation at $x=x_1$ and $x=x_2$. More precisely, from \eqref{rho1new} one easily obtains the trivial result $\rho^{x}(S) = \frac{1}{L^d} \rho(S) $ and from \eqref{rho2new} one obtains
\bea 
&& \int dS_1 dS_2
\rho_W^{x_1,x_2}(S_1,S_2) [ e^{ \lambda_1 S_1 }-1]  [ e^{ \lambda_2 S_2 }-1]    
=    \langle  \tilde u_1(x_1)  \tilde u_2(x_2) \rangle_\xi \nn \, ,
\eea
with $\tilde{u}_i(x_i)$ still given by the solutions of the instanton equation \eqref{inst2} of Sec.~\ref{subsec:analytical:totalsize}, a result that should be compared to \eqref{rho22}. Using the perturbative solution \eqref{u1} in adimensioned units we obtain the Laplace transform of the connected density $\rho_W^{c,x_2-x_1}(S_1,S_2) = \rho_W^{x_2-x_1}(S_1,S_2) - \rho^{x_1}(S_1) \rho^{x_2}(S_2)$ exactly at order $O(\epsilon)$ as
\begin{eqnarray} \label{soluZW} 
\int dS_1 dS_2
\rho_W^{c,x_2-x_1}(S_1,S_2) [ e^{ \lambda_1 S_1 }-1]  [ e^{ \lambda_2 S_2 }-1]  && = - \hat{\Delta}''(W)  \int \frac{d^d q}{(2 \pi)^d} e^{i q (x_1-x_2)}  \frac{Z(\lambda_1)}{1-2Z(\lambda_1) + q^2} \frac{Z(\lambda_2)}{1-2Z(\lambda_2) + q^2 } \nonumber  \, .
\end{eqnarray}

This can be explicitly computed and we obtain the following exact result ($O(\epsilon)$) for this Laplace transform
\begin{eqnarray} \label{ResultLT}
&& \int dS_1 dS_2
\rho_W^{c,x_2-x_1}(S_1,S_2) [ e^{ \lambda_1 S_1 }-1]  [ e^{ \lambda_2 S_2 }-1] =  - \hat \Delta''(W) Z(\lambda_1)Z(\lambda_2) {\cal{I}} ( 1- 2 Z(\lambda_1) ,  1- 2 Z(\lambda_2) , |x_1 - x_2|) \nn \\
&& {\cal{I}} ( a ,  b , x) = \int \frac{d^d q}{(2 \pi)^d} \frac{e^{iqx}}{(a+q^2)(b+q^2)} = \frac{1}{b-a} \left(  \tilde{{\cal{I}}} ( a , x)  - \tilde{{\cal{I}}}( b , x)  \right) \nonumber \\
&&  \tilde{{\cal{I}}} ( a , x) = \frac{1}{(2 \pi)^{d/2}} a^{\frac{d-2}{4}} x^{\frac{2 - d}{2}} K_{ \frac{d-2}{2}}( \sqrt{a}x) \, .
\end{eqnarray}

This is an exact expression but not easy to inverse Laplace transform to obtain the density. For that purpose it is more convenient to work in Fourier space, defining $\rho_W^{q}(S_1,S_2) = \int d^d x e^{i q x} \rho_W^{x}(S_1,S_2)$. We obtain for the connected part
\bea \label{resprhoW}
\rho_W^{c,q}(S_1,S_2) =- \hat{\Delta}''(W) \psi_q(S_1) \psi_q(S_2) \, ,
\eea
where we have introduced the function $\psi_q(S)$ defined as
\bea \label{psiqS} 
\psi_q(S) =\hat{\rho}(S) (1+q^2) \frac{S}{2} \left( 1 - \frac{ \sqrt{\pi S}}{2} q^2 e^{\frac{q^4 S}{4}}   {\rm Erfc}( \frac{q^2 \sqrt{S}}{2})  \right) \, ,
\eea
with $\hat{\rho}(S)$ still given by \eqref{eq:defhatrho}, and whose Laplace transform is
\bea
\int dS (e^{\lambda S} -1) \psi_q(S) = \frac{Z(\lambda)}{1 - 2 Z(\lambda) + q^2} \, .
\eea
We give here a few of the lowest order moments of the function $\psi_q(S)$:
\bea  \label{id}
\int   \psi_q (S) dS = \frac{1}{2}  \quad , \quad \int   \psi_q(S) S dS = \frac{1}{1+q^2}  \quad , \quad \int   \psi_q(S) S^2 dS = \frac{2 ( 3 + q^2)}{(1+q^2)^2}  \, .
\eea
The zeroth order moment allows for example to calculate from \eqref{resprhoW} the connected total density of avalanches
with seeds $x_1$ and $x_2$
\be
\rho_2^c(W,x_2-x_1) := \int dS_1 dS_2 \rho_W^{c,x_2-x_1}(S_1,S_2)  = - \frac{\hat{\Delta}''(W)}{4} \delta^d(x_2-x_1)
\ee 
The form of the result shows that 
the drop in density of avalanche at $w=W$ caused by the occurence of an avalanche at $w=0$ with seed
$x_1$ is only felt at microscopic distances. This is not so surprising 
since the vast majority of avalanches are of microscopic sizes (the density $\rho^{x}(S)$ is not normalizable at small $S$). 
By contrast, the first joint moment (obtained from the second identity in \eqref{id})
is dominated by large avalanches and reads
\bea
\langle S_1 S_2 \rangle_{\rho_W^{c,x}}  
:= \int dS_1 dS_2 S_1 S_2 \rho_W^{c,x}(S_1,S_2)
=- \hat{\Delta}''(W) 2^{-1-d/2} \pi^{-d/2} x^{2-\frac{d}{2}} 
K_{\frac{d-4}{2}}(x) \, .
\eea
Integrating this result over $x$ leads to the exact relation $L^{-2 d} \langle S_1 S_2 \rangle_{W}^c =  - \frac{\hat{\Delta}''(W)}{L^d}$. This suggests that the above result may be quite accurate since it integrates to an exact result. Suprisingly, the full $x$ dependence of this first joint moment is
{\it identical} to the one of the correlation of the local sizes when the driving is uniform, see Eq.~112 in 
\cite{ThieryStaticCorrelations}. This surprising result does not hold for higher order moments as can easily be seen by comparing the formulas.

\subsubsection{Universality and the massless limit}
\label{sec:univmassless}

We now address in more depth the issue of universality and of the massless limit. 
In Section \ref{sec:calculation1} we obtained a universal scaling form \eqref{scalform}
for $\rho^c_W(S_1,S_2)$. This form however involves two features which
makes it not fully universal. (i) First it depends explicitly on $m$ and $S_m$, i.e. it is universal 
(i.e. independent of small scale details) within the model of the driving parabolic well,
which provides a large scale cutoff. One can ask if it is possible to obtain
results for avalanche correlations which would be independent of the details of large scale cutoff (which we call full universality).
(ii) The second feature is that $\rho^c_W(S_1,S_2)$ is proportional to $(m L)^d$, i.e. the number of independent
regions along the interface. This makes sense since avalanches are expected to
be correlated only if they are separated within a distance $x \sim 1/m$ along the interface.

In the above calculation of $\rho^{c,x}_W(S_1,S_2)$ the factor $(m L)^d$ is absent, 
since the separation of the seeds, $x$, is fixed. In Fourier space we now expect, restoring units
\bea \label{qscal} 
\rho_W^{c,q}(S_1,S_2) = \frac{1}{m^d S_m^4} {\cal F}_d(\frac{q}{m},\frac{W}{W_m},\frac{S_1}{S_m},\frac{S_2}{S_m})
\eea
where ${\cal F}_d$ is a universal function. From our result \eqref{resprhoW} we see that, restoring units,
with $w=W/W_m$ one has the expansion around the upper critical dimension
\bea \label{Fd12} 
{\cal F}_d(q,w,s_1,s_2)
= - m^{d-4} \Delta''(W) \psi_{q}(s_1) \psi_{q}(s_2) + O(\epsilon^2) 
= - A_d \tilde \Delta^{* \prime \prime}(w) 
\psi_{q}(s_1) \psi_{q}(s_2) + O(\epsilon^2) 
\eea 
where all factors are universal using the fixed point \eqref{scalingdelta}
(apart for a single non-universal scale, see discussion below \eqref{scaluniform}). 

Let us now discuss the behavior of \eqref{qscal} in the region $q \gg m$, $S_i \ll S_m$, $W \ll W_m$
where we hope to find a fully universal behavior. It is equivalent to take $m \to 0$ and 
consider {\it the massless limit}. It turns out that consideration of this limit, upon some hypothesis,
leads a host of information, i.e. determination of some critical exponents.\\

Let us first recall the analysis of the massless limit for the single avalanche size density, $\rho(S)$, and its connection
to the Narayan-Fisher conjecture \cite{NarayanDSFisher1992b,NarayanDSFisher1993a}. \\

The massless limit of the starting equation of motion for the interface, Eq. \eqref{eq:model:intro},
is obtained by defining $f(x,t) = m^2 w(x,t)$, the applied force. One can then take $m \to 0$ at fixed
$f(x,t)$ and the equation remains well defined. Then one must define densities per unit force, 
denoted everywhere with a subscript $f$, rather than
per unit $w$ as we did until now. They simply differ by the factor $m^2$, e.g.
\be
\rho^f(S) = m^{-2} \rho(S)
\ee
It is easy to see on the result for the BFM \eqref{eq:defhatrho} that the massless limit
of $\rho^f(S)$ is well defined (i.e. all factors of $m$ cancel) leading to
\be
\rho^f(S)= \lim_{m \to 0} m^{-2} \rho(S) = \frac{L^d}{\sigma^{1/2}} \frac1{2 \sqrt{\pi} S^{3/2}} 
\ee
which is the fully universal part of the density corresponding to small avalanches $S \ll S_m$.
It turns out that this extends beyond mean field. Indeed, for any $d$ one has, via simple dimensional analysis
\bea
\rho(S) = \frac{L^d}{S_m^2} r(S/S_m) \quad , \quad r(s) \sim_{s \to 0} s^{-\tau} 
\eea
where $\tau$ is the avalanche size exponent. For $\rho^f(S)= \lim_{m \to 0} m^{-2} \rho(S)$ to
be finite, we see that we need $m^{-2} S_m^{\tau-2}$ to be finite in the limit $m \to 0$, and 
using $S_m \sim m^{-(d+\zeta)}$ this is equivalent to
\be
\tau = \tau_{\rm NF} = 2 - \frac{2}{d+\zeta} 
\ee
This relation was tested to one loop in \cite{LeDoussalWiese2012a}, and is rather natural since we do expect
that a universal massless limit exists (a massless field theory). We will thus generally assume
that densities per unit force do exist in the limit $m \to 0$. This led in \cite{DobrinevskiLeDoussalWiese2014a} to predictions
for a number of other avalanche exponents, and we call it the {\it generalized NF conjecture}. \\

Consider now the massless limit of the joint size density to which we apply similar arguments. We note 
from \eqref{psiqS} that
\bea
&& \lim_{m \to 0} m^4 \psi_{q/m}(\frac{S}{S_m}) = \frac{\sigma}{S} \psi(q (S/\sigma)^{1/4}) \\
&& \psi(q) = \frac{q^2}{4 \sqrt{\pi }}-\frac{1}{8}
   e^{\frac{q^4}{4}} q^4
   \text{erfc}\left(\frac{q^2}{2}\right) = \frac{q^2}{4 \sqrt{\pi}} + O(q^4) = \frac{1}{2 \sqrt{\pi }
   q^2}+O\left(\frac{1}{q^4}\right) \label{psiq}
\eea
where $\psi(q)$ is the massless scaling form of $\psi_q(S)$. Thus our result to $O(\epsilon)$ for the joint density per unit force has a well defined $m \to 0$ limit
\bea \label{eps1} 
\rho_W^{c,q,f}(S_1,S_2) = \lim_{m \to 0} m^{-4} \rho_W^{c,q}(S_1,S_2) 
=  - \frac{\Delta''(W)}{\sigma^2} \frac{1}{S_1 S_2} \psi(q (S_1/\sigma)^{1/4}) 
\psi(q (S_2/\sigma)^{1/4})  + O(\epsilon^2) 
\eea 
Note that as $m \to 0$, $\Delta''(W) = A_d m^\epsilon \tilde \Delta^{* \prime \prime}(W m^\zeta) \simeq 
A_d \tilde \Delta^{* \prime \prime}(0^+) + O(\epsilon^2)$ if $W$ is kept fixed as $m \to 0$. Hence the
dependence in $W$ disappears in that limit since there can be avalanches of arbitrary sizes, all
avalanches can be considered as successive (i.e. $W=0^+$).\\

More generally, we surmise that this massless limit exist in any dimension. Scaling arguments and dimensional analysis then lead to the scaling form \footnote{Note that an additional dependence in $q W^{1/\zeta}$ cannot be ruled out, although it does not seem to appear to $O(\epsilon)$ (see remark above). Thus, to be fully consistent,
in \eqref{fulluniv} we have in mind here $W=0^+$, i.e. successive avalanches.}
\bea \label{fulluniv} 
\rho_W^{c,q,f}(S_1,S_2) = \frac{\ell_\sigma^{d+4}}{(S_1S_2)^{\tau_c^1} }
~ f_d( \ell_\sigma \, q S_1^{1/(d+\zeta)}, 
\ell_\sigma \, q S_2^{1/(d+\zeta)}) \quad , \quad \tau_c^1 = \frac{1}{2}(2 - \frac{4-d-2\zeta}{d+\zeta})
\eea 
where $\tau_c^1$ is a correlation exponent, $f_d$ is a fully universal function and the scale $\ell_\sigma$ is non fully universal 
\footnote{$\ell_\sigma$ has dimension $(x^\zeta/u)^{1/(d+\zeta)}$ where $x$ and $u$ are
lengths in internal and displacement directions respectively.}. For the parabolic well model
it equals $\ell_\sigma = \lim_{m \to 0} m^{-1} S_m^{-1/(d+\zeta)}$ (hence can be measured
independently using \eqref{defSm}), as easily seen by studying the possible $m \to 0$ limit of the scaling function 
${\cal F}_d$ in \eqref{qscal}. \\

We see that for $d=4-\epsilon$ and $\zeta=O(\epsilon)$ the form \eqref{fulluniv} reproduces \eqref{eps1} with
$\ell_\sigma=\sigma^{-1/4}$ and
\bea
f_d(q_1,q_2) = - A_d \tilde \Delta^{* \prime \prime}(0^+) \psi(q_1) \psi(q_2) + O(\epsilon^2) 
= - \frac{16 \pi^2 \epsilon}{9}  \psi(q_1) \psi(q_2) + O(\epsilon^2)
\eea 
where $\psi(q)$ is given in \eqref{psiq}.\\

Let us now study the $q=0$ limit, i.e. the uniform driving studied in Section \ref{sec:calculation1}. For a fixed $m>0$ 
one has
\bea
\rho_W^c(S_1,S_2) = L^d \rho_W^{c,q=0}(S_1,S_2) = - \frac{L^d}{m^d S_m^4} m^{d-4} \Delta''(W)
\frac{S_1 S_2}{4 S_m^2} \hat \rho(S_1/S_m) \hat \rho(S_2/S_m) 
\eea 
using \eqref{qscal}, \eqref{Fd12},
and, from \eqref{psiqS}, that $\psi_{q=0}(s)= \hat \rho(s) \frac{s}{2}$. It coincides with 
\eqref{rhocW} upon using \eqref{eq:defhatrho}. Its $m \to 0$ limit reads
\bea
\rho_W^{c,f}(S_1,S_2) = m^{-4} \rho_W^c(S_1,S_2) = - (m L)^d  
\frac{A_4 \tilde \Delta^{* \prime \prime}(0^+)}{\sigma^3} 
\frac1{16 \pi S_1^{1/2} S_2^{1/2}} + O(\epsilon^2) \, ,
\eea
Note that this result cannot be obtained from taking the $q \to 0$ limit of \eqref{eps1}
since $\psi(q) \sim q^2$ as $q \to 0$. Hence there is a non-commutation of limits
$m \to 0$ and $q \to 0$. \\

It is reasonable to surmise that in any dimension, as $m \to 0$ \footnote{one cannot
exclude an additional factor $g_d(S_1/S_2)$ (not present to this order) which we ignore here for simplicity
(it does not affect the discussion of the critical exponent $\tau_c$ defined here for $S_1 \sim S_2$).}
\bea
\rho_W^{c,f}(S_1,S_2) \sim (m L)^d \frac{1}{S_1^{\tau_c} S_2^{\tau_c} }  
\eea 
with $\tau_c=1/2$ in mean-field, i.e. for $d=d_c=4$ here, the factor $(m L)^d$ being the number
of independent regions. One can obtain this factor by considering the $q \to 0$ limit
of the massless result on one hand, and the $q=0$ massive result at small $m$ on the other, 
and requiring matching upon setting $q=m$. This determines the $q \to 0$ behavior of the scaling 
function $f_d$ as 
\bea
f_d( q S_1^{1/(d+\zeta)},
q S_2^{1/(d+\zeta)}) \simeq_{q \to 0} q^d (S_1 S_2)^{\frac{1}{2} \frac{d}{d+\zeta}} 
\eea 
So that, substituting $q=m$ in \eqref{fulluniv} we indeed obtain
\bea \label{tauc} 
\rho_W^{c,f}(S_1,S_2) = L^d \rho_W^{c,q=m,f}(S_1,S_2)
\sim (L m)^d \frac{1}{S_1^{\tau_c} S_2^{\tau_c} }  \quad , \quad \tau_c = 2 - \frac{2+d}{d+\zeta} 
\eea 
where the exponent $\tau_c$ is thus fully determined, via this generalized NF argument
(recovering $\tau_c=1/2$ in mean-field for $d=4$, $\zeta=0$).

%
%\bigskip
%*** to be erased ***
%
%Note the limit $x << 1/m$ which leads to $\rho_W^x(S_1,S_2) = 
%\rho_{x_1}(S_1) \rho_{x_2}(S_2) (1- \Delta''(W) |x|^{4-d})$ plus smooth.
%
%
%

\section{Numerics} \label{sec:numerics}

In this section we compare some of our results with the simulation of a $d=1$ elastic interface with short-ranged elasticity in a short-ranged correlated disordered landscape.

\subsection{Protocol}

To perform numerical simuations we choose a Gaussian disorder $F(u,x)$ with a correlator $\overline{F(u,x)F(u',x')} = \delta(x-x') \Delta_0(u-u')$ with $\Delta_0(u) = \sigma \delta u e^{-|u|/\delta u}$ with $\delta u$ the microscopic correlation length of the disorder. As explained in \cite{ThieryShape}, this can be realized by taking $F(u,x)$ as a collection (indexed by $x$) of independent Ornstein-Uhlenbeck processes (in the direction $u$). More precisely the model we study can be defined as, for any driving protocol $w(x,t)$
\bea
&& \eta \partial_t u(x,t) = \nabla^2_x u(x,t) - m^2( u(x,t)-w(x,t) ) + F(u(x,t),x)  \nn \\
&& \partial_u F(u,x) = \sqrt{2 \sigma} \xi(u,x) - \frac{1}{\delta u} F(u,x) \, .
\eea
 with $\xi(u,x)$ a unit centered two dimensional Gaussian white noise $\overline{\xi(u,x)\xi(u',x')} = \delta(u-u') \delta(x-x')$. The advantage of this setup is that one can directly obtain an autonomous equation for the velocity field $\dot{u}(x,t)$: the above model is equivalent to
\bea \label{Simu}
&& \eta \dot{u}(x,t) = \nabla^2 \dot{u}(x,t) + m^2(\dot{w}(t)-\dot{u}(x,t)) + \partial_t {\sf F}(x,t)  \ , \nn \\
&& \partial_t {\sf F}(x,t) = \sqrt{2 \sigma \dot{u}(x,t)} \chi(x,t) - \frac{\dot{u}(x,t)}{ \delta u} {\sf F}(x,t) \ ,
\eea
with $\chi(x,t)$ a centered Gaussian white noise $\overline{\chi(x,t) \chi(x',t')} = \delta^d(x-x') \delta(t-t')$ and we have the equality in law ${\sf F}(x,t) \sim F(x,u(x,t))$. 

We take as initial conditions $\dot{u}(x,t=0)={\sf F}(x,t=0)=0$ and then apply a sequence of kicks of size $\delta w =1$, which amounts at setting $\dot{u}(x,t) = \frac{m^2 }{\eta } \delta w$ at the beginning of each kick and wait for the interface to stop before applying the next kick. The motion of the interface between each kick is measured by integrating the velocity field in between two kicks and this defines for each kick an avalanche $S_x$. The avalanche at the $n$-th kick is said to have been triggered at $w = n \delta w$. We wait for the sytem to reach a stationary state before measuring anything. Averages are obtained using $50$ independent `experience', each experience consisting in $2\times10^6$ kicks, and we have thus simulated $10^8$ avalanches. Correlations between avalanches are measured for avalanches inside a window of $1000$ successive kicks.

In the results reported here we have taken an interface of lateral extension $L=1024$ with periodic boundary conditions, discretized with $1024$ points. The parameters are chosen as $m =20/L \simeq 0.02$. The kicks are of size $\delta w = 1$ and the microscopic disorder correlation length is taken as $\delta u = 5 \delta w=5$. The discretization in time is handled using an algorithm similar as the one introduced in \cite{Dornic} and we take a time step $\delta t = 0.025$.

\subsection{Results}

\subsubsection{Renormalized disorder correlator}

Central to our results is the measurement of the renormalized disorder second cumulant $\Delta(w-w') = L^d m^4 \overline{(u(w)-w) (u(w')-w')}^c$ where $u(w)$ is the position of the interface in the end of the $w/\delta w$-th kick. The plot of $\Delta(w)$ is presented in Fig.~\ref{fig:Delta}. To obtain a good measurement of the derivative $\Delta'(w)$ and $\Delta''(w)$ we fitted $\Delta(w)$ with a polynomial of order seven and differentiated directly the fitted polynomial. A plot of $\Delta'(w)$ and $\Delta''(w)$ is also given in Fig.~\ref{fig:DeltaPS}.

\begin{figure}
\centering
\includegraphics[width=7cm]{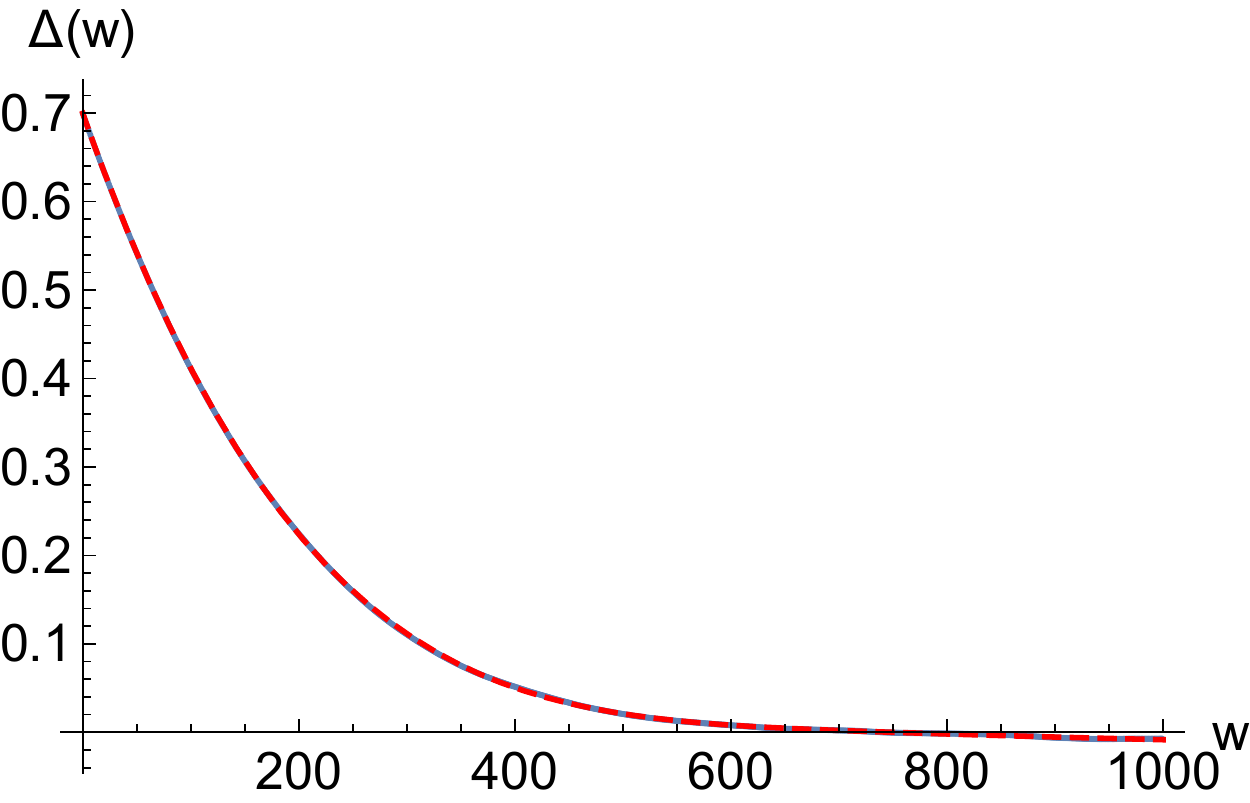}
\caption{The renormalized disorder second cumulant measured in the simulations (blue dots) and its polynomial fit (red line).}
\label{fig:Delta}
\end{figure}

\begin{figure}
\centering
\includegraphics[width=7cm]{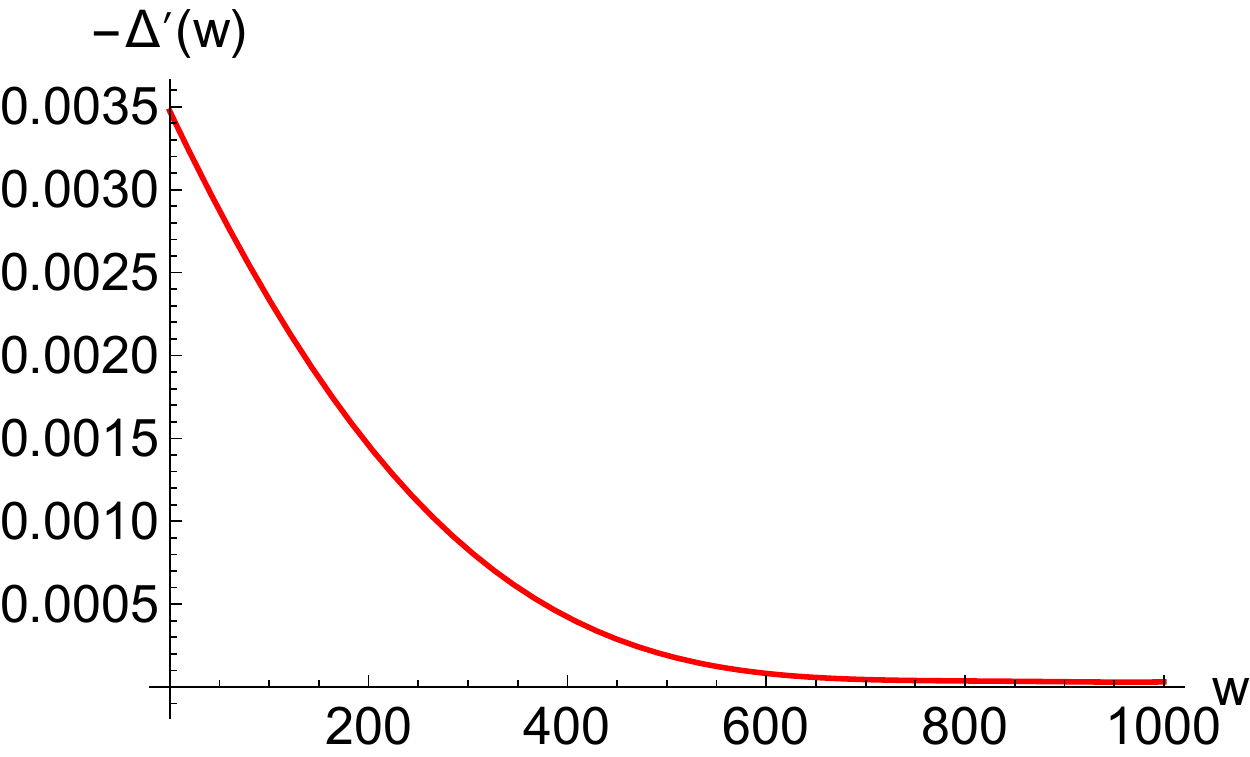} \includegraphics[width=7cm]{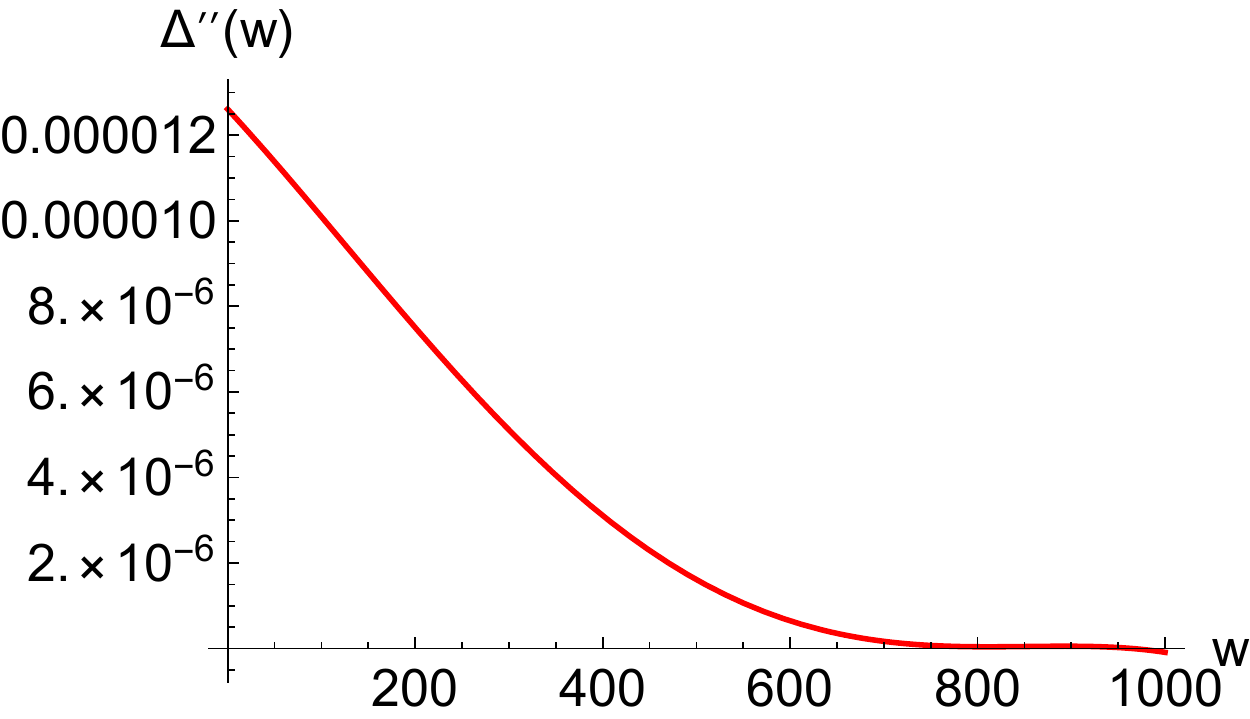}
\caption{First (left figure) and second (right figure) derivative of the renormalized disorder second cumulant as obtained using the polynomial fit of the measured renormalized disorder second cumulant.}
\label{fig:DeltaPS}
\end{figure}

\subsubsection{Total avalanche sizes correlations}

We show in Fig.~\ref{fig:S1S2global} the comparison between the measurements of $\langle S_1 S_2 \rangle_W^c$ and $\frac{1}{2}\left( \langle S_1 S_2^2 \rangle_W^c +\langle S_1^2 S_2 \rangle_W^c\right)$ and our predictions \eqref{eq:resS1S2}-\eqref{eq:resS1S22}. The analytical result for $\langle S_1 S_2 \rangle_W^c$ is exact and the agreement is as expected perfect. The analytical result for $\frac{1}{2}\left( \langle S_1 S_2^2 \rangle_W^c +\langle S_1^2 S_2 \rangle_W^c\right)$  is only a $O(\epsilon)$ approximation and appears to overestimate the correlations. 

\begin{figure}
\centering
\includegraphics[width=7cm]{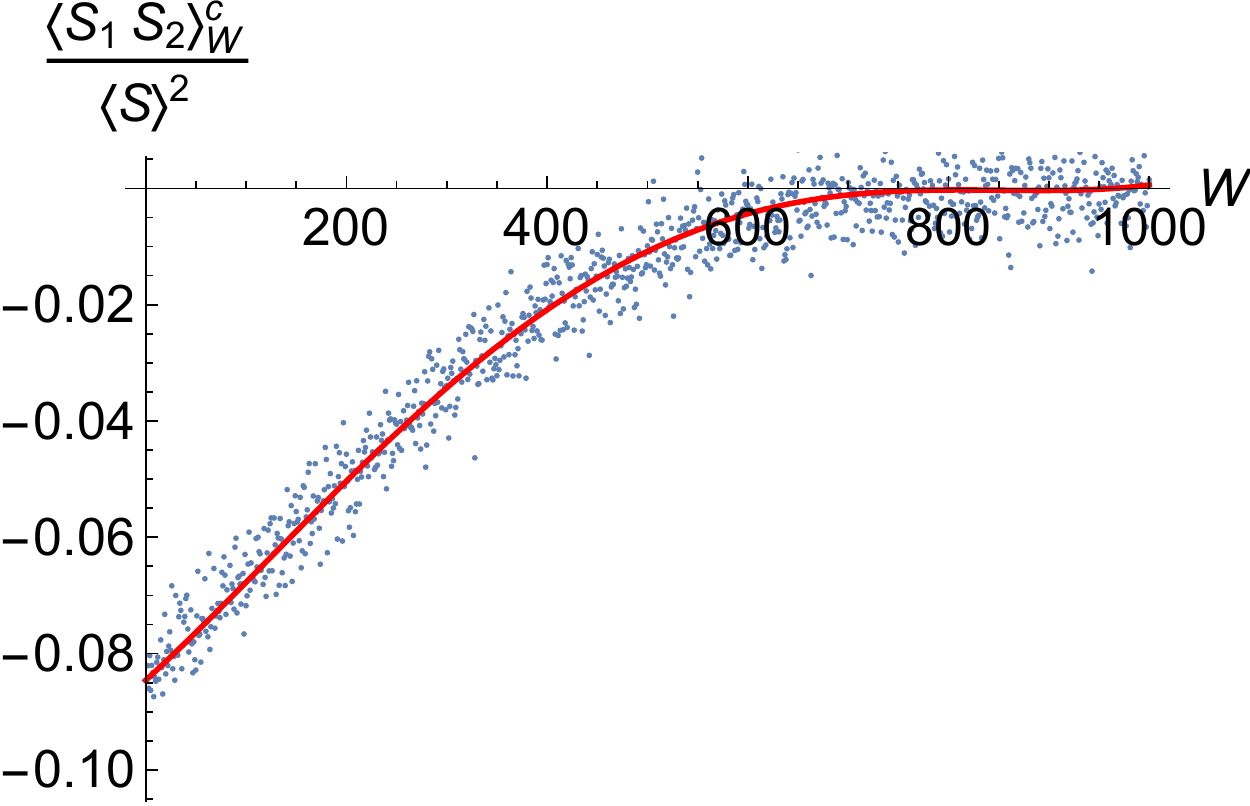} \includegraphics[width=7cm]{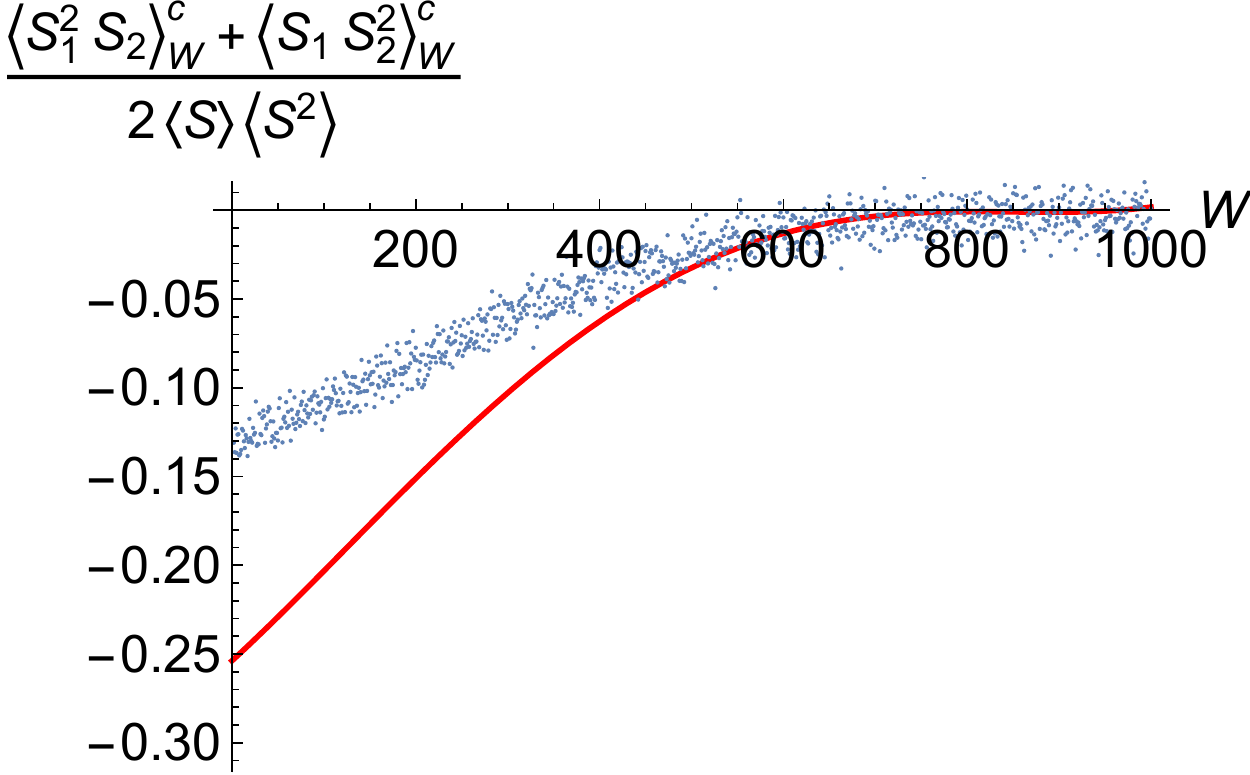}
\caption{Left: Comparison between the measurement of  $\langle S_1 S_2 \rangle_W^c$ (blue dots) and our prediction \eqref{eq:resS1S2} (exact result).  Right: Comparison between the measurement of  $\frac{1}{2}\left( \langle S_1 S_2^2 \rangle_W^c +\langle S_1^2 S_2 \rangle_W^c\right)$ (blue dots) and our prediction \eqref{eq:resS1S22} (order $O(\epsilon)$ result). The blue dots corresponds to direct measurements of the correlations between avalanches, each dot corresponding to an average over avalanches for a given $W$. The dispersion of the cloud of dots gives an estimate of the accuracy of the measurement.}
\label{fig:S1S2global}
\end{figure}

\subsubsection{Local avalanche sizes correlations}

We show in Fig.~\ref{fig:S1S2local} the comparison between the measurements of $\langle S_{10} S_{2x} \rangle_W^c$ and our prediction \eqref{eq:resS1x1S2x2}. Despite the fact that \eqref{eq:resS1x1S2x2} is only valid up to order $O(\epsilon)$ (with here $\epsilon=3$), it is clear that it is a very good approximation. Also it seems that our result tends to slightly underestimate the correlations between avalanches at short distance and overestimate the correlations at large distance.

\begin{figure}
\centering
\includegraphics[width=8cm]{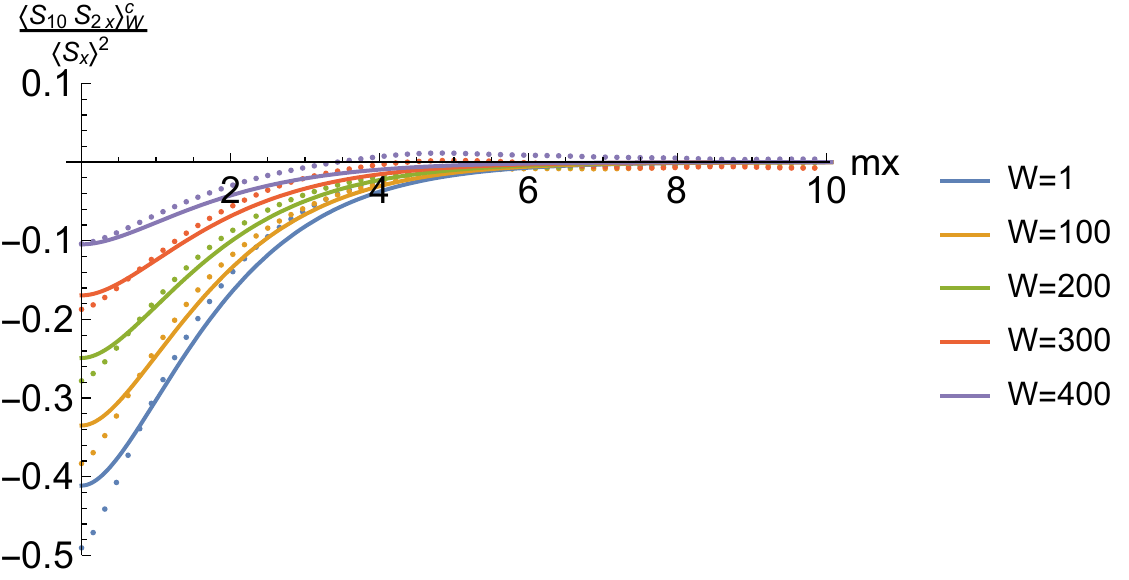} \includegraphics[width=8cm]{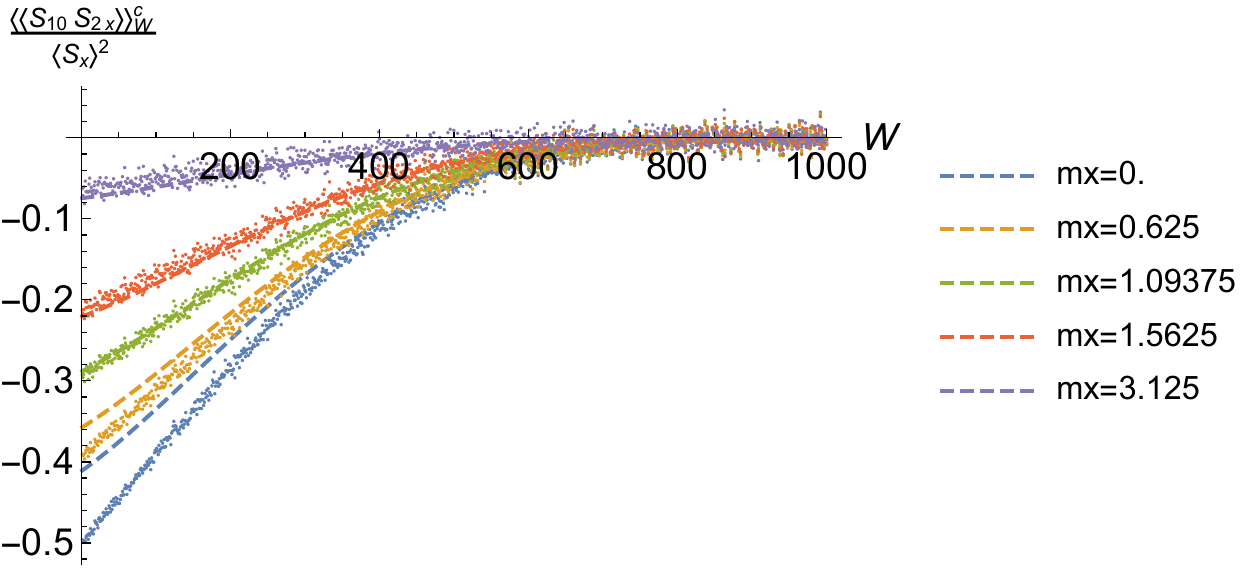}
\caption{Comparison between the measurement of  $\langle S_{10} S_{2x} \rangle_W^c$ (dots) and our prediction \eqref{eq:resS1x1S2x2} (plain and dashed lines, order $O(\epsilon)$ approximation) as a function of $mx$ for a few values of $W$ (left) or as a function of $W$ for a few values of $mx$ (right). }
\label{fig:S1S2local}
\end{figure}

\section{Analytical results: Dynamical correlations}
\label{sec:dynamical} 

\subsection{Correlations of the total velocities in an avalanche, and of the avalanche durations } 
\label{subsec:corrvel} 

In this Section we study the correlation of the global velocities (center of mass
velocities) in the two avalanches. As a by product we also obtain the correlation
between the avalanche durations. Let us define the total (areal) velocities in the two avalanches,
which we denote $\dot u_1 \equiv \dot u_1(t_1) =  \int d^dx \dot u_1(x,t_1)$ and 
$\dot u_1 \equiv \dot u_2(t_2) =  \int d^dx \dot u_1(x,t_2)$. The time is 
counted from the kick in each avalanche, i.e. each avalanche starts at $t_i=0$. \\

Similar methods as in Section \ref{subsec:analytical:localsize} give, for the joint density $\rho^{x_1,x_2}_W(\dot u_1,\dot u_2)$ at fixed positions of the seeds $x_1,x_2$
\bea \label{rho23}
&& \int d\dot u_1 d\dot u_2
\rho^{x_1,x_2}_W(\dot u_1,\dot u_2) [ e^{ \lambda_1 \dot u_1 }-1]  
[ e^{ \lambda_2 \dot u_2 }-1]   
= \langle \tilde{u}_1(x_1,0) \tilde{u}_2(x_2,0) \rangle_\xi 
\eea
where $\tilde u_i(x,t)$ are the solutions of the time-dependent space inhomogeneous instanton equation 
\bea \label{inst30} 
( \partial_t +   \nabla_x^2 - 1) \tilde u_i(x,t) + \tilde u_i(x,t)^2 = -\xi_i(x) \tilde u_i(x,t)
 - \lambda_i \delta(t-t_i) 
\eea 
with $\tilde u_i(x,t>t_i)=0$, which are needed only to first order in each
$\xi_i$. As in Section \ref{subsec:analytical:localsize}  we introduce
\bea
\tilde u_i(x,t)  = \tilde u^0_i(t) +  \tilde u^1_i(x,t)  
\eea 
where $\tilde u^0_i(t)$ is the solution for $\xi_i(x)=0$, which satisfies
\bea \label{inst31} 
( \partial_t  - 1) \tilde u^0_i(t) + \tilde u^0_i(t)^2 = - \lambda_i \delta(t-t_i) 
\eea 
with $\tilde u^0_i(t>t_i)=0$. The solution is well known to be \cite{LeDoussalWiese2011a}
\be
\tilde u_i^0(t) = \frac{\lambda_i}{\lambda_i + (1-\lambda_i) e^{t_i-t} } \theta(t<t_i)
\ee
Let us recall that from this solution one obtains
the single avalanche (time-dependent) density in the BFM, denoted $\rho(\dot u) \equiv \rho_t(\dot u)$, 
of the total velocity $\dot u(t)=\int d^d x \dot u(x,t)$,
by Laplace inversion of
\bea
\int_0^{+\infty} d\dot u \rho_t(\dot u) (e^{\lambda \dot u}-1) = L^d \tilde u^0(0)  = L^d \frac{\lambda}{\lambda + (1-\lambda) e^{t} } 
\quad \Rightarrow \quad \rho_t(\dot u) = L^d \frac{e^{t-\frac{e^t \dot u}{e^t-1}}}{\left(e^t-1\right)^2} \quad , \quad 
\dot u>0 
\eea 
By integration over time one recovers the well known result 
for the 
mean density of total velocity $\rho_{v=0^+} (\dot u)$ 
for a uniform driving in the limit $v=0^+$, 
$\rho_{v=0^+} (\dot u) = \int_0^{+\infty} dt \rho_t(\dot u) = \frac{L^d}{\dot u} e^{- \dot u}$
(see \cite{LeDoussalWiese2011a,LeDoussalWiese2012a,ABBMNonstat2012}). 
%{\red add reference 
%to "Nonstationary dynamics of the ABBM model.." noted DLW below}).
The same calculation also gives the density of the {\it avalanche duration} $T$
in the BFM \footnote{Indeed one also has that $\int_0^{+\infty} d\dot u \rho_t(\dot u)= \frac{1}{e^t-1}
= \partial_{\delta w=0^+} (1-p_{\delta w}(t))$, where $1-p_{\delta w}(t)$ is the probability that
the velocity is non-zero at time $t$. This comes from the
definition of the density $\rho(\dot u) = \partial_{\delta w=0^+} P_{\delta w}(\dot u)$ 
from the PDF of the total velocity, which
reads $P_{\delta w}(\dot u)=p_{\delta w}(t) \delta(\dot u) + (1-p_{\delta w}(t)) \tilde P_{\delta w}(\dot u)$
where $\tilde P_{\delta w}$ is the smooth normalized PDF for $\dot u>0$
(see (26) and (28) in \cite{ABBMNonstat2012} for exact expressions). Note the
extra delta function piece which is usually not considered
in the expression for the density.}
\be \label{durationrho} 
\rho(T) = - \partial_t \int_0^{+\infty} d\dot u \rho_t(\dot u) |_{t=T}
= \frac{L^d}{4 \sinh^2(T/2)}
\ee 
which reads, in dimensionfull units, $\rho(T) = \frac{L^d}{S_m \tau_m 4 \sinh^2(T/2\tau_m)}$.
\\

To obtain the connected joint density, we need to calculate 
$\tilde u^1_i(x,t)$ to first order in $\xi_i$. For that purpose
we introduce the dressed response kernel 
\bea  \label{defpropag2}
 ( \partial_t + \nabla_x^2 - 1 + 2 \tilde u_i^0(t) ) G_i(x,t;y,t') = - \delta^d(x-y) \delta(t-t')  \, .
 \eea 
with $G_i(x,t;y,t')=0$ for $t>t'$ (i.e. time is in effect reversed as compared to a
standard response function). It reads in Fourier, for $t,t'<t_i$
\bea
G_i(q,t,t') &=& e^{-(q^2+1)(t'-t) + 2 \int_{t}^{t'} \tilde u_i^0(s) ds} \theta(t'-t) \\
&=& e^{-(q^2+1)(t'-t)} \frac{(1-\lambda_i + \lambda_i e^{t'-t_i})^2}{(1-\lambda_i + \lambda_i e^{t-t_i})^2}
\theta(t'-t) \nonumber
\eea
We obtain
\bea
 \tilde u^1_i(x,t)  = \int d^dy \, G_i(x,t;y,t') \xi_i(y) \tilde u_i^0(t') + O(\xi^2) 
\eea 
which leads to the Laplace transform of the connected density as
\bea \label{rho2udot}
&& \int d\dot u_1 d\dot u_2
\rho^{c,x_1,x_2}_W(\dot u_1,\dot u_2) [ e^{ \lambda_1 \dot u_1 }-1]  
[ e^{ \lambda_2 \dot u_2 }-1]   \\
&& = - \hat{\Delta}''(W)
\int d^dy \int_0^{t_1} dt' \int_0^{t_2} dt'' G_1(x_1,0;y,t') G_2(x_2,0;y,t'') \tilde u^0_1(t') \tilde u^0_2(t'') \, . \nonumber
\eea
It is more convenient to work in Fourier space and define
\bea
\rho^{c,x_1,x_2}_W(\dot u_1,\dot u_2) = \int \frac{d^d q}{(2 \pi)^d} e^{- i q (x_1-x_2)} 
\rho^{c,q}_W(\dot u_1,\dot u_2)
\eea
%Note that in effect the time integrals are on a finite interval.
%\bea \label{rho2udot}
%&& \int d\dot u_1 d\dot u_2
%\rho^{c,q}_W(\dot u_1,\dot u_2) [ e^{ \lambda_1 \dot u_1 }-1]  
%[ e^{ \lambda_2 \dot u_2 }-1]   \\
%&& = - \hat{\Delta}''(W)
%\int_0^{t_1} dt' \int_0^{t_2} dt'' G_1(q,0,t') G_2(q,0,t'') \tilde u^0_1(t') \tilde u^0_2(t'') \, . 
%\eea
% obtain
We finally obtain the Laplace transform of the connected part of the joint density of total velocities in the two avalanches,
for a fixed driving wavevector $q$ as
\bea \label{rho2udot2}
&& \int d\dot u_1 d\dot u_2
\rho^{c,q}_W(\dot u_1,\dot u_2) [ e^{ \lambda_1 \dot u_1 }-1]  
[ e^{ \lambda_2 \dot u_2 }-1]  = - \hat{\Delta}''(W) F_{\lambda_1}(q,t_1) F_{\lambda_2}(q,t_2) 
\eea
where we have defined
\bea
F_{\lambda}(q,t) =  \lambda \frac{ \left(1-q^2-\lambda\right) e^{-(1+q^2) t}
    + (1-\lambda) \left(q^2-1\right) e^{-t}+\lambda  q^2 e^{-2 t}}{q^2
   \left(q^2-1\right) \left(1- \lambda + \lambda e^{-t} \right)^2}
\eea 

%{\tiny
%\bea
%F_{\lambda}(q,t) = 
%\frac{ \lambda  e^{(2-q^2) t} \left(2-\lambda
%   -q^2\right)-\lambda  \left(\lambda -\lambda  q^2+(\lambda -1)
%   \left(q^2-2\right) e^t\right)}{\left(q^4-3 q^2+2\right) \left(\lambda
%   -(\lambda -1) e^t\right)^2}
%\eea}
We now analyze this formula in various cases: (i) homogeneous driving
(ii) fixed distance between the seeds (iii) massless limit (leading to
conjectures for the correlation exponents in any dimension).

\subsubsection{Homogeneous driving}

{\bf Velocities}. For the homogeneous driving $\delta w(x)=\delta w$, the Laplace transform of the connected joint density
simplifies into
\bea \label{rho2udot2}
&& \int d\dot u_1 d\dot u_2
\rho^c_W(\dot u_1,\dot u_2) [ e^{ \lambda_1 \dot u_1 }-1]  
[ e^{ \lambda_2 \dot u_2 }-1]  = - L^d \hat{\Delta}''(W) F_{\lambda_1}(t_1) F_{\lambda_2}(t_2) 
\eea
where we denote
%\bea
%&& F_{\lambda}(t) = F_{\lambda}(q=0,t) =
%-\frac{\lambda  \left(\lambda  \left(e^t (t-1)+1\right)-e^t t\right)}{\left(\lambda
%   -(\lambda -1) e^t\right)^2} 
%\eea
\bea
&& F_{\lambda}(t) = F_{\lambda}(q=0,t) =
-\frac{\lambda  e^{-t} \left(\lambda  \left(t-1+e^{-t}\right)- t\right)}{\left(1-\lambda+\lambda e^{-t}\right)^2} 
\eea
%{\tiny
%\bea
%&& F_{\lambda}(t) = F_{\lambda}(q=0,t) = -\frac{\lambda  \left(e^t-1\right) \left((\lambda -2) e^t-\lambda \right)}{2
%   \left(\lambda -(\lambda -1) e^t\right)^2} 
%\eea }

It is possible to perform the inverse Laplace transform explicitly and obtain
\bea
&& \rho^c_W(\dot u_1,\dot u_2) =  - L^d \hat{\Delta}''(W) ~ r_{t_1}(\dot u_1) \, r_{t_2}(\dot u_2) \\
&& r_t(\dot u) = \frac{e^{t-\frac{e^t u}{e^t-1}} \left(e^t \left(-(t+1) \dot u+e^t   (t+\dot u-2)+4\right)-t-2\right)}{\left(e^t-1\right)^4} =  \frac{d}{dt} \frac{\left(-t+e^t-1\right) e^{t-\frac{e^t \dot u}{e^t-1}}}{\left(e^t-1\right)^2}
\eea 
Restoring the units it reads
\bea \label{rhocuu} 
\rho^c_W(\dot u_1,\dot u_2)
=
- \frac{L^d}{m^4 v_m^2 S_m^2} \Delta''(W)  \, r_{t_1/\tau_m}(\frac{\dot u_1}{v_m}) r_{t_2/\tau_m}(\frac{\dot u_2}{v_m})
\eea
where $v_m =S_m/\tau_m$. One can check that
$\int_0^{+\infty} dt_1 \int_0^{+\infty} dt_2 \langle \dot u_1 \dot u_2 \rangle_W^c$
calculated with this formula coincides with the result for 
$\langle S_1 S_2 \rangle_W^c$ obtained above in \eqref{eq:resS1S2} (which, we recall
was an exact result, i.e. valid beyond the $\epsilon$ expansion). 
 \\

We can calculate the joint density of the mean total velocity, averaged over all the avalanche.
In dimensionless units, using that $\int_0^{+\infty} dt \, r_t(\dot u) = e^{-\dot u}$, it reads simply
\be \label{integrated} 
 \int_0^{+\infty} dt_1 \int_0^{+\infty} dt_2  \rho^c_W(\dot u_1,\dot u_2) =  - L^d \hat{\Delta}''(W) 
 e^{-\dot u_1-\dot u_2}
 \ee
Note that it is regular at small $\dot u$, unlike the single avalanche density (see above)
$\rho(\dot u)=\frac{1}{\dot u} e^{-\dot u}$. \\

Let us obtain some cumulants. One has (in dimensionfull units)
\be
\frac{\langle \dot u_1 \dot u_2 \rangle^c_W}{\langle \dot u_1 \rangle \langle \dot u_2 \rangle} = - \frac{\Delta''(W)}{m^4 L^d} \frac{t_1  t_2}{\tau_m^2}
\ee
and (in dimensionless units)
\be
\frac{\langle \dot u_1 \dot u_2^3 \rangle^c_W}{\langle \dot u_1^2 \dot u_2^2 \rangle^c_W} =
\frac{3 e^{t_1-t_2} \left(e^{t_2}-1\right) \left(-3
   t_2+e^{t_2} \left(t_2+2\right)-2\right)}{2 \left(-2
   t_1+e^{t_1} \left(t_1+1\right)-1\right) \left(-2
   t_2+e^{t_2} \left(t_2+1\right)-1\right)}
\ee\\
 
{\bf Durations}. Finally we can obtain the correlation between the durations $T_1$ and $T_2$ of the two avalanches.
Integrating over $\dot u$ 
 \be
 \int_0^{+\infty} d\dot u \, r_t(\dot u)= \frac{e^t (t-1)+1}{\left(e^t-1\right)^2}
 \ee
 hence we obtain \footnote{Again, one shows that 
 $\int_0^{+\infty} \int_0^{+\infty}  d\dot u_1 d\dot u_2 \rho_W(\dot u_1,\dot u_2)
 = \partial_{\delta w_1}|_{\delta w_1=0} \partial_{\delta w_2}|_{\delta w_2=0} 
 {\rm Prob}(\dot u_1>0,\dot u_2>0) =  \rho_W(T_1,T_2)$ since
 $ {\rm Prob}(\dot u_1>0,\dot u_2>0)= {\rm Prob}(T_1>t_1,T_2>t_2)$.}
 \bea \label{dT} 
&&  \rho^c_W(T_1,T_2) =  - L^d \hat{\Delta}''(W)  d(T_1) d(T_2) \\
&&  d(T) = - \partial_T  \int_0^{+\infty} d\dot u \, r_T(\dot u) =
  \frac{e^T \left(e^T (T-2)+T+2\right)}{\left(e^T-1\right)^3} \simeq_{T \ll 1} \frac{1}{6}-\frac{T^2}{60}+O\left(T^3\right) \\
&& ~~~~~~~~~~~~~~~~~~~~~~~~~~~~~~~~~~~~~~~~~~~~~~~~~~~~~~~~~~~~~~~~~~~~~~ \simeq_{T \gg 1} \simeq T e^{-T}
 \eea
where $d(T)$ is a decreasing function of $T$. Note that the dimensionful version is
\be \label{durunits} 
  \rho^c_W(T_1,T_2) =  - \frac{L^d }{m^4  S_m^2 \tau_m^2} \Delta''(W)  d(T_1/\tau_m) d(T_2/\tau_m) 
\ee
\\

\subsubsection{Fixed distance between the seeds.}

{\bf Velocities}. We now calculate the connected joint density 
$\rho^{c,x}_W(\dot u_1,\dot u_2)$, for the total 
velocities (at times $t_1$ and $t_2$ respectively) 
of two avalanches starting a distance $x$ apart (i.e. in $x_1$ and $x_2$ with $x_2-x_1=x$).
The definitions are similar to those in Section \ref{subsec:seedcentered}.
One finds
\bea
&& \rho^{c,x}_W(\dot u_1,\dot u_2) = \int \frac{d^d q}{(2 \pi)^d} \, e^{i q x } \, \rho^{c,q}_W(\dot u_1,\dot u_2) \\
&& \rho^{c,q}_W(\dot u_1,\dot u_2) =  -  \hat{\Delta}''(W) ~ r_{q,t_1}(\dot u_1) 
r_{q,t_2}(\dot u_2) 
\eea  
where
\bea
&& r_{q,t}(\dot u) = \frac{e^{-q^2 t -\frac{e^t \dot u}{e^t-1}+t}}{q^2 \left(q^2-1\right) \left(e^t-1\right)^4}
(A(q,t) + \dot u B(q,t)) \\
&& B(q,t) = e^t (- q^2 e^t +e^{q^2 t}+q^2-1) \\
&& A(q,t) = -2 q^2 e^t-2 q^2 e^{q^2 t+t}+ (q^2-1) (1+e^{(q^2+2)t}) +(q^2+1) ( e^{2 t} + e^{q^2 t}) 
\eea

Restoring the units it reads
\bea \label{veloc} 
 \rho^{c,q}_W(\dot u_1,\dot u_2) =
- \frac{1}{m^4 v_m^2 S_m^2} \Delta''(W)  \, r_{q/m,t_1/\tau_m}(\frac{\dot u_1}{v_m}) \, r_{q/m,t_2/\tau_m}(\frac{\dot u_1}{v_m})
\eea
where $v_m =S_m/\tau_m$. Again one checks that 
$\int_0^{+\infty} dt_1 \int_0^{+\infty} dt_2 \langle \dot u_1 \dot u_2 \rangle_W^{c,q}$
calculated with this formula coincides with the result for 
$\langle S_1 S_2 \rangle_W^{c,q}$ obtained in Section \ref{sec:calc}.\\

{\bf Durations}. We can obtain the joint duration density. In dimensionless units, using
\be \int_{0}^{+\infty} d \dot u \, r_{q,t}(\dot u) =
\frac{(q^2-1) e^t-q^2 + e^{(1-q^2) t} }{q^2
   \left(q^2-1\right) \left(e^t-1\right)^2}
\ee
we find
\bea \label{dq} 
&& \rho^{c,q}_W(T_1,T_2) =  -  \hat{\Delta}''(W)  \, d_q(T_1) \, d_q(T_2) \\
 && d_q(t)=- \partial_t \int_{0}^{+\infty} d \dot u \, r_{q,t}(\dot u)  = \frac{e^t \left(\left(q^2-1\right) e^t+e^{-q^2 t} \left(\left(q^2+1\right)
   e^t-q^2+1\right)-q^2-1\right)}{q^2 \left(q^2-1\right) \left(e^t-1\right)^3}
\eea
%
%\bea
%&& r_{q,t}(\dot u) = 
%\frac{e^{-q^2 t -\frac{e^t \dot u}{e^t-1}+t}}{\left(q^4-3 q^2+2\right) \left(e^t-1\right)^4}
%(A(q,t) + \dot u B(q,t)) \\
%&& B(q,t) = e^t \left(\left(q^2-2\right) e^t+e^{q^2 t}-\left(q^2-1\right) e^{2 t}\right) \\
%&& A(q,t) = \left(e^t-1\right) \left(e^t \left(q^2 \left(e^t-1\right)+2\right)+e^{q^2 t}
%   \left(\left(q^2-2\right) e^t-q^2\right)\right)
%\eea

\subsubsection{Massless limit}

We now study these formula in the massless limit, to extract the fully universal limit. 
We follow the same strategy as explained in Section \ref{sec:univmassless}. \\

{\bf Velocities}. In the limit $m \to 0$, we obtain from \eqref{veloc}
\bea \label{rfudot} 
\rho^{c,q,f}_W(\dot u_1,\dot u_2) = \lim_{m \to 0}  m^{-4} \rho^{c,q}_W(\dot u_1,\dot u_2) =  -  \Delta''(W) 
\frac{1}{t_1} \tilde r_{q \sqrt{t_1}}(\frac{\dot u_1}{t_1})
\frac{1}{t_2} \tilde r_{q \sqrt{t_2}}(\frac{\dot u_2}{t_2}) + O(\epsilon^2)
\eea 
in the massless units (i.e. such that $\sigma=\eta=1$), where
\bea \tilde r_{q}(\dot u)  = \lim_{m \to 0} m^2 r_{q/m,m^2}(m^2 \dot u) =
\frac{e^{-q^2-\dot u} \left(q^2+2- (q^2+1) \dot u+e^{q^2}
   (q^2+\dot u-2) \right)}{q^4} &=& \frac{1}{2} \dot u e^{-\dot u} + O(q^2) \label{commute} \\
                                      & = & \frac{e^{- \dot u}}{q^2} + O(q^{-4}) 
\eea
We recall that $\Delta''(W) \simeq A_d \tilde \Delta^{* \prime \prime}(0^+) + O(\epsilon^2)$ 
if $W$ is kept fixed as $m \to 0$.

We can thus surmise, more generally in the limit $m \to 0$ (and fixed $W$), 
from scaling and dimensional analysis, the fully
universal scaling form \footnote{up to two non universal scales $\ell_\sigma= \lim_{m \to 0} m^{-1} S_m^{-1/(d+\zeta)}$ and
$\ell_\eta= \lim_{m \to 0} m^{-1} \tau_m^{-1/z}$.}
\bea
\rho^{c,q,f}_W(\dot u_1,\dot u_2) = \frac{1}{(t_1 t_2)^{{\sf a}_c^1}} %\frac{1}{z} (\frac{3}{2} d - 2 + 2 \zeta)-1}}
~ F_d(\frac{\dot u_1}{t_1^{\frac{d+\zeta}{z}-1}}, \frac{\dot u_2}{t_2^{\frac{d+\zeta}{z}-1}}, q t_1^{1/z}, q t_2^{1/z}) 
\quad , \quad {\sf a}_c^1 = \frac{1}{z} (\frac{3}{2} d - 2 + 2 \zeta)-1
\eea 
with, in the $d=4-\epsilon$ expansion
\bea
F_d(\dot u_1,\dot u_2,q_1,q_2) = - A_4 \tilde \Delta^{* \prime \prime}(0^+) \tilde r_{q_1}(\dot u_1) 
\tilde r_{q_2}(\dot u_2) + O(\epsilon^2) 
\eea 
with $A_4 \tilde \Delta^{* \prime \prime}(0^+)= \frac{16 \pi^2}{9} \epsilon$. \\

Let us study $q=0$, i.e. the homogeneous driving. From \eqref{rhocuu} using that
$r_{m^2 t}(m^2 \dot u) \simeq_{m \to 0} \frac{\dot u}{2 t^2 m^2}  e^{- \dot u/t}$ we obtain the
simple and finite expression in the massless limit 
\bea \label{rfudot} 
\rho^{c,f}_W(\dot u_1,\dot u_2) = \lim_{m \to 0}  m^{-4} \rho^c_W(\dot u_1,\dot u_2) =  - L^d \Delta''(W) \frac{\dot u_1}{2 t_1^2}  e^{- \dot u_1/t_1} 
\frac{\dot u_2}{2 t_2^2}  e^{- \dot u_2/t_2} 
\eea 
in the massless units. At variance with the joint size densities, there is no factor $(m L)^d$ (although of
course there is a factor $L^d$). Hence this expression already has a fully universal limit. The 
origin of this surprising fact is that now there is a commutation of limits $q \to 0$ and $m \to 0$, as
can be seen from \eqref{commute}. Presumably it occurs because the times $t_1,t_2$ provide some
natural cutoff
\footnote{Note however, interestingly, the non-commutation of limits $m \to 0$ and
integration over time, as integrating the massless result \eqref{rfudot} over time gives an extra factor $1/4$
as compared to taking the small mass (or velocity) limit of the formula \eqref{integrated} 
which was integrated over time at finite $m$. This is
because the time scale $\tau_m$ diverges in that limit, while \eqref{rfudot} is dominated
by time scales $t_1,t_2$.}. If we surmise that for this observable this property holds more generally
we obtain
\bea
\rho^{c,f}_W(\dot u_1,\dot u_2) = \frac{L^d}{(t_1 t_2)^{\frac{1}{z} (\frac{3}{2} d - 2 + 2 \zeta)-1}}
F_d(\frac{\dot u_1}{t_1^{\frac{d+\zeta}{z}-1}}, \frac{\dot u_2}{t_2^{\frac{d+\zeta}{z}-1}},0,0) 
\eea 
 
{\bf Durations}.
Let us now discuss the joint density of the avalanche durations. Restoring the units in \eqref{dq} we have
\bea \label{dq2} 
&& \rho^{c,q}_W(T_1,T_2) =  -  \frac{1}{m^4  S_m^2 \tau_m^2}  \Delta''(W)  \, d_{q/m}(T_1/\tau_m) \, d_{q/m}(T_2/\tau_m)
\eea
In the limit $m\to 0$ we obtain
\bea \label{dq3} 
&& \rho^{c,q,f}_W(T_1,T_2) =  \lim_{m \to 0} m^{-4}  \rho^{c,q}_W(T_1,T_2) =
- \Delta''(W)  \frac{1}{T_1 T_2} \, \tilde d(q T_1^{1/2})  \, \tilde d(q T_2^{1/2}) 
\eea
in massless units with
\bea
\tilde d(q) = \frac{q^2-2 + e^{-q^2} (q^2+2)}{q^4} = \frac{q^2}{6} + O(q^4) = \frac{1}{q^2} + O(\frac{1}{q^4}) 
\eea

We can compare with the massless limit of \eqref{durunits} 
\be \label{durunits2} 
\rho^{c,f}_W(T_1,T_2) \simeq_{m \to 0} 
 - (m L)^d m^{4-d} \Delta''(W) \frac{1}{36} + O(\epsilon^2) 
 \ee
in massless units. As was the case for the size joint density there is a factor $(m L)^d$ and 
non commutation of limits $q \to 0$ and $m \to 0$. The matching, i.e. setting $q=m$ and $m \to 0$
into \eqref{dq3} and recovering \eqref{durunits2} (to $O(\epsilon)$)
also works, as was the case for the size density. Another sign of the non commuting
limits is that if one integrates the $q=0$ result \eqref{rfudot} over $\dot u_1,\dot u_2$ 
and take $\partial_{t_1} \partial_{t_2}$ one obtains zero, while the correct
subleading term in $m$ is  \eqref{durunits2}.\\

More generally we can thus surmise, from scaling and dimensional analysis, the fully
universal scaling form (up to two non universal scales) in the massless limit in 
general dimension $d$
\bea \label{TT} 
\rho^{c,q,f}_W(T_1,T_2) = \frac{1}{(T_1 T_2)^{\alpha^1_c}}
G_d(q T_1^{1/z}, q T_2^{1/z}) \quad , \quad \alpha^1_c = 1 - \frac{1}{2 z} (4-d - 2 \zeta)
\eea 
with, in the $d=4-\epsilon$ expansion, $G_d(q_1,q_2) = - A_4 \tilde \Delta^{* \prime \prime}(0^+)  \tilde d(q_1) \tilde d(q_2) + O(\epsilon^2)$, with $A_4 \tilde \Delta^{* \prime \prime}(0^+)= \frac{16 \pi^2}{9} \epsilon$. By the same reasoning which led to \eqref{tauc} we can also surmise that
the r.h.s of \eqref{TT} must behave as $q^d$ at small $q$ leading to
\bea
\rho_W^{c,f}(T_1,T_2) \sim (m L)^d \frac{1}{T_1^{\alpha_c} T_2^{\alpha_c} }  \quad , \quad \alpha_c = 
1 - \frac{1}{z} (2- \zeta)
\eea 
with $\alpha_c=0$ in mean-field, i.e. for $d=d_c=4$, consistent with \eqref{durunits2}.

\subsection{Correlation of the shapes of two avalanches}
\label{subsec:shape} 

Consider now the joint density of the total velocities 
$\dot u_i = \dot u_i(t_i) = \int d^d x \, \dot u_i(x,t_i)$, $i=1,2$
and $\dot u'_i = \dot u_i(t'_i) = \int d^d x \, \dot u_i(x,t'_i)$, $i=1,2$
in two avalanches, the times $0< t_1<t_1'$ and $0< t_2<t_2'$ being counted from the beginning of each
avalanche. Its Laplace transform satisfies
\bea \label{rho234}
&& \int d\dot u_1 d\dot u'_1 d\dot u_2 d\dot u'_2
\rho^{x_1,x_2}_W(\dot u_1,\dot u'_1,\dot u_2,\dot u'_2) [ e^{ \lambda_1 \dot u_1 + \lambda'_1 \dot u'_1  }-1]  
[ e^{ \lambda_2 \dot u_2 + \lambda'_2 \dot u'_2 }-1]   
= \langle \tilde{u}_1(x_1,0) \tilde{u}_2(x_2,0) \rangle_\xi 
\eea
where now $\tilde{u}_i(x,t)$ are solution of \eqref{inst30} with the source
$- \lambda_i \delta(t-t_i)  - \lambda'_i \delta(t-t'_i)$ with $t_i < t'_i$. \\

Here we are only interested in the correlation of the shape of each avalanche. Let us first recall the definition of the mean shape for a single avalanche, at fixed avalanche duration $T$ : it is the mean
velocity as a function of time, conditioned to the avalanche duration
\be \label{shapeBFM1} 
\langle \dot u_i(t_i) \rangle_{T_i} = \frac{\int d\dot u_i \dot u_i \rho(\dot u_i, T_i) }{\rho(T_i)} 
\ee
where $\rho(\dot u_i, T_i)$ is the joint density of the velocity and duration in an avalanche.
Here we are interested in the correlation of the shapes 
\be \label{shapeseeds} 
\langle \dot u_1(t_1) \dot u_2(t_2) \rangle^{x_1,x_2}_{T_1,T_2} 
= \frac{\int d\dot u_1 d\dot u_2 \dot u_1 \dot u_2 \rho^{x_1 x_2}_W(\dot u_1, T_1,\dot u_2, T_2) }{\rho^{x_1 x_2}_W(T_1,T_2)} 
\ee
with fixed positions of the seeds at $x_1,x_2$, and in the correlation of the shapes 
for a homogeneous driving
\be \label{shapeuniform} 
\langle \dot u_1(t_1) \dot u_2(t_2) \rangle_{T_1,T_2} 
= \frac{\int d\dot u_1 d\dot u_2 \dot u_1 \dot u_2 \rho_W(\dot u_1, T_1,\dot u_2, T_2) }{\rho_W(T_1,T_2)} 
\ee

The denominator, i.e. the joint density of durations $\rho^{x_1 x_2}_W(T_1,T_2)$ was studied in the previous section. The numerator can be obtained by taking a derivative of \eqref{rho234} w.r.t. $\lambda_1$ and $\lambda_2$ at 
$\lambda_1=\lambda_2=0$, and setting $\lambda_i'=-\infty$ (which implies $\dot u_i(t'_i)=0$ hence $T_i<t'_i$)
and, to obtain the joint density with durations $T_i$, taking the derivative w.r.t. $t'_i$
\be \label{corru1u2} 
\int d\dot u_1 d\dot u_2 \dot u_1 \dot u_2 \rho^{x_1 x_2}_W(\dot u_1, T_1,\dot u_2, T_2)
= \partial_{t'_1}|_{t'_1=T_1} \partial_{t'_2}|_{t'_2=T_2}
\lim_{\lambda'_1,\lambda'_2 \to - \infty} 
\partial_{\lambda_1}|_{\lambda_1=0} \partial_{\lambda_2}|_{\lambda_2=0} \langle \tilde{u}_1(x_1,0) \tilde{u}_2(x_2,0) \rangle_\xi 
\ee

Let us first describe the solution of \eqref{inst30} to order 0 in $\xi$ and recall the calculation of the 
shape in the BFM. The solution of 
\bea \label{inst31b} 
( \partial_t  - 1) \tilde u^0_i(t) + \tilde u^0_i(t)^2 = - \lambda_i \delta(t-t_i)  - \lambda'_i \delta(t-t'_i)
\eea 
with $\tilde u_i^0(t>t'_i)=0$ is \cite{LeDoussalWiese2012a,ABBMNonstat2012}
\bea
&& \tilde u_i^0(t) = \frac{\lambda'_i}{\lambda'_i + (1-\lambda'_i) e^{t'_i-t} } \quad , \quad t_i<t<t'_i   \\
&& \tilde u_i^0(t) = \left(1- \frac{\lambda_i \lambda'_i e^{t_i} - (1-\lambda_i)(1-\lambda'_i) e^{t'_i}}{
(1+ \lambda_i) \lambda'_i e^{t_i} + (1- \lambda'_i) \lambda_i e^{t'_i}} e^{t_i-t} \right)^{-1} \quad , \quad t < t_i 
\eea
It allows to obtain the mean shape of a single avalanche \eqref{shapeBFM1} within the BFM as
\be \label{shapeBFM2} 
\langle \dot u_i(t_i) \rangle_{T_i} =  \frac{1}{\rho^{x_i}(T_i)} \partial_{t'_i}|_{t'_i=T_i} \partial_{\lambda_i}|_{\lambda_i=0} \, \tilde{u}_i^0(0)|_{\lambda'_i=-\infty} 
\ee
Thus, for the remainder of the calculation we only need $\tilde{u}_i^0(0)$ for $\lambda_i'=-\infty$ and to first order in $\lambda_i$, which reads
\bea \label{trunc} 
\tilde u_i^0(t) = \frac{\theta(t'_i-t)}{1 - e^{t'_i-t}} \left(1 + \lambda_i \frac{e^{- t_i} (e^{t_i}-e^{t'_i})^2}{e^{t}-e^{t'_i}} \theta(t_i-t) \right)
\eea 
where we discard higher order terms in $\lambda_i$. This leads to the classical BFM result for the
shape \cite{LeDoussalWiese2012a,ABBMNonstat2012}
\bea \label{shape0} 
\langle \dot u_i(t_i) \rangle_{T_i} &=&   \frac{1}{\rho^{x_i}(T_i)} \partial_{t'_i}|_{t'_i=T_i}
\frac{e^{- t_i} (e^{t_i}-e^{t'_i})^2}{(1-e^{t'_i})^2} = \frac{2 (1- e^{-t_i})(e^{T_i}-e^{t_i})}{e^{T_i}-1} \\
&= & s_0(t_i,T_i)=
\frac{4 \sinh \frac{t_i}{2} \sinh \frac{T_i-t_i}{2} }{\sinh \frac{T_i}{2} } = 2 T_i z(1-z) + O(T^3) \quad , \quad z=\frac{t_i}{T_i}
\nonumber
\eea
using \eqref{durationrho}. Restoring the units the BFM result reads
\be \label{bfmshape} 
\langle \dot u(t) \rangle_{T} = v_m \, s_0(\frac{t}{\tau_m}, \frac{T}{\tau_m}) \simeq 
\frac{v_m}{\tau_m} \, T \, 2 z(1-z) + O(T^3) 
\ee
with $v_m = S_m/\tau_m$, which has a well defined massless limit since in the BFM $\lim_{m \to 0} \frac{v_m}{\tau_m}=\sigma/\eta^2=1$ in dimensionless units. Note that
the next order term $O(T^3)$ is $\sim 1/\tau_m^2 \sim m^4$ higher order in that limit. 
\\

Now we study \eqref{inst30} to the desired order $O(\xi)$. As in Section \ref{subsec:corrvel} we obtain
\bea
\tilde u_i(x,t)  = \tilde u^0_i(t) +  \int d^dy \, G_i(x,t;y,t') \xi_i(y) \tilde u_i^0(t') + O(\xi^2) 
\eea 
Inserting in \eqref{corru1u2} and going to Fourier space, the connected correlation
\bea
\rho^{c,x_1,x_2}_W(\dot u_1,T_1,\dot u_2,T_2) = \int \frac{d^d q}{(2 \pi)^d} e^{- i q (x_1-x_2)} 
\rho^{c,q}_W(\dot u_1,T_1,\dot u_2,T_2)
\eea
becomes (taking into account that the disconnected piece has been substracted) 
\bea \label{aa} 
&& \int d\dot u_1 d\dot u_2 \dot u_1 \dot u_2 \rho^{c,q}_W(\dot u_1, T_1,\dot u_2, T_2)  = - \hat{\Delta}''(W) \, \hat r_{q,t_1}(T_1) \hat r_{q,t_2}(T_2) \\
&& \hat r_{q,t_i}(T_i) =   \partial_{t'_i}|_{t'_i=T_i} \partial_{\lambda_i}|_{\lambda_i=0} \int_0^{t'_i} dt'  G_i(q,0,t') \tilde u^0_i(t') \nonumber
\eea 
where we must insert \eqref{trunc} and the propagator to the needed order
\bea
G_i(q,t,t') &=& e^{-(q^2+1)(t'-t) + 2 \int_{t}^{t'} \tilde u_i^0(s) ds} \theta(t'-t) \\
&=& e^{- (q^2+1) (t'-t) } \frac{(e^{t_i'} - e^{t'})^2}{(e^{t_i'} - e^{t})^2}
\theta(t<t'<t_i') \bigg( 1 + \lambda_1 \frac{2 e^{- t_i} (e^{t_i} - e^{t'_i})^2 (e^t-e^{t'})}{(e^t - e^{t_i'}) (e^{t'_i}-e^{t'})}
\theta(t<t'<t_i) \nonumber \\
&& 
+ \lambda_1 \frac{2 e^{- t_i} (e^{t_i} - e^{t}) (e^{t_i}-e^{t'_i})}{e^t - e^{t_i'} }
\theta(t<t_i<t') \bigg) \nonumber 
\eea\\

From this one obtains the connected shape correlation. Since the connected parts of the densities 
are $O(\epsilon)$ one can expand \eqref{shapeseeds} as
\bea \label{connshapedef} 
&& \langle \dot u_1(t_1) \dot u_2(t_2) \rangle^{c, x_1,x_2}_{T_1,T_2} = 
\langle \dot u_1(t_1) \dot u_2(t_2) \rangle^{x_1,x_2}_{T_1,T_2} - \langle \dot u_1(t_1) \rangle_{T_1} \langle \dot u_2(t_2) \rangle_{T_2} \\
&& = \frac{\int d\dot u_1 d\dot u_2 \dot u_1 \dot u_2 \rho^{c,x_1 x_2}_W(\dot u_1, T_1,\dot u_2, T_2) }{L^{-2d} \rho(T_1) \rho(T_2)} - \langle \dot u_1(t_1) \rangle_{T_1} \langle \dot u_2(t_2) \rangle_{T_2} 
\frac{\rho^{c,x_1 x_2}_W(T_1, T_2) }{L^{-2d} \rho(T_1) \rho(T_2)} + O(\epsilon^2) \nonumber 
\eea
We recall that the non-connected parts are $x_1,x_2$ independent and $\rho^x(T)=L^{-d} \rho(T)$.
Thus, to this order, one can easily Fourier transform and write the shape correlation at fixed seed positions as
\bea
&& \langle \dot u_1(t_1) \dot u_2(t_2) \rangle^{c, x_1,x_2}_{T_1,T_2} =
\int \frac{d^d q}{(2 \pi)^d} e^{- i q (x_1-x_2)} \langle \dot u_1(t_1) \dot u_2(t_2) \rangle^{c,q}_{T_1,T_2} \\
&& 
\langle \dot u_1(t_1) \dot u_2(t_2) \rangle^{c,q}_{T_1,T_2} = - \hat{\Delta}''(W) [ \frac{\hat r_{q,t_1}(T_1) \hat r_{q,t_2}(T_2)}{L^{-2d}  \rho(T_1) \rho(T_2)} 
- \frac{d_{q}(T_1) d_{q}(T_2)}{L^{-2d}  \rho(T_1) \rho(T_2)} \langle \dot u_1(t_1) \rangle_{T_1} 
\langle \dot u_1(t_2) \rangle_{T_2}  ] + O(\epsilon^2) 
\eea
where we have used \eqref{aa}. We recall that $\rho(T)$ is given in \eqref{durationrho} and $d_q(T)$ in \eqref{dq}.
For the shape correlation at uniform driving we obtain
\bea \label{shapeu} 
&& \langle \dot u_1(t_1) \dot u_2(t_2) \rangle^{c}_{T_1,T_2} =
- L^d \hat{\Delta}''(W) [ \frac{\hat r_{t_1}(T_1) \hat r_{t_2}(T_2)}{\rho(T_1) \rho(T_2)} 
- \frac{d(T_1) d(T_2)}{\rho(T_1) \rho(T_2)} \langle \dot u_1(t_1) \rangle_{T_1} 
\langle \dot u_1(t_2) \rangle_{T_2}  ] + O(\epsilon^2)
\eea
where $\hat r_t(T)=\hat r_{q=0,t}(T)$ and $d(T)=d_{q=0}(T)$ is given in \eqref{dT}.
In both formula, for $\langle \dot u(t) \rangle_{T}$ one can insert to this order the BFM shape 
given in \eqref{shape0}.\\

Here we will only discuss the final formula for the shape at homogeneous driving, i.e. for $q=0$.
The formula at finite $q$ are given in the Appendix \ref{app:formula}.\\

Denoting $z=t/T$, we obtain the building blocks of \eqref{shapeu} as
 \bea
A_T(t)= \frac{\hat r_{t}(T)}{L^{-d} \rho(T)} &=&  \frac{\sinh(T-t)+(t-2 T) \cosh (t)-(t+T) \cosh (t-T)+\sinh(t)+2 T
-\sinh (T)+T\cosh (T)}{\sinh^2 \frac{T}{2}} \nonumber \\
& =& %\frac{1}{3} T^3 \left(z^4-2  z^3+z\right)+O\left(T^5\right) =  
\frac{1}{3} T^3 z (1-z) \left(1+z (1-z) \right) +O\left(T^5\right)  
\eea
and
\bea
&& B_T(t) = \frac{d(T)}{L^{-d} \rho(T)} \langle \dot u(t) \rangle_T = \frac{ 4 \sinh (\frac{t}{2}) \sinh(\frac{T-t}{2})  }{\sinh (\frac{T}{2})} \left(T \coth
   (\frac{T}{2})-2\right)
   = \frac{1}{3} T^3 z (1-z) +O\left(T^5\right)
\eea 
From them one obtains the explicit expression for \eqref{shapeu} in the form, restoring units
\be \label{shapeu2} 
 \langle \dot u_1(t_1) \dot u_2(t_2) \rangle^{c}_{T_1,T_2} =
- (m L)^{-d} v_m^2 m^{d-4} \Delta''(W) \left( A_{T_1/\tau_m}(\frac{t_1}{\tau_m}) 
A_{T_2/\tau_m}(\frac{t_2}{\tau_m}) - B_{T_1/\tau_m}(\frac{t_1}{\tau_m}) 
B_{T_2/\tau_m}(\frac{t_2}{\tau_m}) \right) + O(\epsilon^2)
\ee
with $v_m=S_m/\tau_m$.
Note that since $A_T(T-t)=A_T(t)$ and $B_T(T-t)=B_T(t)$ we find that to this
order the shape correlations are
symmetric in independently changing each $t_i \to T_i-t_i$. As is well known this
property of the mean shape for a single avalanche does not hold to the
next order in $\epsilon$, the corrections having been obtained in \cite{DobrinevskiLeDoussalWiese2014a}. \\

We now display the shape correlation explicitly for small avalanches. This is equivalent to
consider the small mass limit. Putting together
the above results we find in the small $T_1,T_2$ limit, with $z_1=t_1/T_1$ and $z_2=t_2/T_2$
\be
 \langle \dot u_1(t_1) \dot u_2(t_2) \rangle_{T_1,T_2}^c = - (m L)^{-d}  \frac{S_m^2}{\tau_m^{8}} m^{d-4} \Delta''(W) 
 \times \frac{1}{9} T_1^3 T_2^3 z_1 (1-z_1) z_2 (1-z_2) \big[ (1+z_1 (1-z_1)) (1+z_2 (1-z_2)) - 1 \big] 
\ee
The factor $(m L)^{-d}$ is expected since, in order to be correlated, the avalanches should take place in
the same region of size $1/m$ along the interface. We see that there is indeed a correlation
between the shapes of the avalanches. However it is of order $O(T_1^3 T_2^3)$, i.e.
loosely speaking it arises as a correlation between the subleading $O(T^3)$ terms in the avalanche shape
in \eqref{bfmshape}. As a consequence it is $O(m^4)$ in the limit of small $m$. 
Hence the correlation of the fully universal part, which corresponds to the parabolic form
for the mean shape, vanishes, but there is a non-trivial correlation at the
next leading order. \\

Finally, performing the double integral $\int_0^{T_1} dt_1 \int_0^{T_2} dt_2$ on \eqref{shapeu2} 
we obtain an interesting observable, the correlation between the total sizes of two avalanches, at fixed durations, 
$\langle S_1 S_2 \rangle_{T_1,T_2}^c$, which reads
\bea \label{shapeu2} 
 \langle S_1 S_2 \rangle_{T_1,T_2}^c &=&
- (m L)^{-d} S_m^2 m^{d-4} \Delta''(W) \left( A(T_1/\tau_m) A(T_2/\tau_m) - B(T_1/\tau_m) B(T_2/\tau_m) \right) + O(\epsilon^2) \\
& = & - (m L)^{-d} \frac{S_m^2}{\tau_m^8} m^{d-4} \Delta''(W) \frac{11}{8100} T_1^4 T_2^4  + O(T_1^6,T_2^6)
\eea
with
\bea
A(T) = T \frac{T (\cosh
   (T)+2)-3 \sinh (T)}{\sinh^2 (\frac{T}{2})} = \frac{T^4}{15} + O(T^6) \quad , \quad B(T) 
   = 2 (T \coth (\frac{T}{2})-2)^2 = \frac{T^4}{18} + O(T^6) 
\eea 
We note the sum rule (in dimensionless units)
\be
\int_0^{+\infty} dT A(T) \rho(T) = \int_0^{+\infty} dT A(T) \frac{1}{4 \sinh^2(T/2)}=1
\ee 
Hence multiplying \eqref{shapeu2} by $\rho(T_1) \rho(T_2)$, the first term
leads exactly to \eqref{eq:resS1S2} (with $\langle S \rangle=L^d$). The second
term comes from the $O(\epsilon)$ correlation between $T_1$ and $T_2$ together
with the precise definition of the connected shape in \eqref{connshapedef}.
Similarly there is a sum rule when performing the double integral $\int_0^{T_1} dt_1 \int_0^{T_2} dt_2$
on Eq. \eqref{aa}. Indeed, upon further integration 
$\int_0^{+\infty} dT_1 dT_2$, it should give back \eqref{resprhoW}. Using \eqref{id} we see
that it implies the sum rule (in dimensionless units)
\be \label{sumrule1} 
\int_0^{+\infty} dT \int_0^{T} dt \, \hat r_{q,t}(T)= \frac{1}{1+ q^2} 
\ee
which we have checked is indeed obeyed by the result in Appendix \ref{app:formula}.

%
%{\red more could be done:}
%
%{\red obtain $\langle S^p_1 S^q_2 \rangle_{T_1,T_2}^c$ and $\langle T^p_1 T^q_2 \rangle_{S_1,S_2}^c$
%and useful check of above} 
%
%{\red analyze the shape as a function of $q$, universality, guess critical exponents.} 
%
%{\red solve the question of the UV divergence upon integration of $\rho^{c,q}$} 

\section{Conclusion and discussion}

In conclusion we have obtained a method to calculate the correlations 
between successive avalanches in the dynamics of an elastic interface near the depinning transition, to leading order in the $\epsilon=d_c-d$ expansion.
This approach is technically simpler than the corresponding one developed to study shocks in
the statics. We have first calculated correlations of the global and local sizes, which, to 
the accuracy of $O(\epsilon)$ leads to results formally similar to the one for the
shocks in the statics, apart from the fact 
that the renormalized disorder correlator is different
in each case. Next we have calculated an observable which is more natural in
the dynamics, the correlation between avalanche sizes with prescribed positions
of the seeds (the starting points). The massless limit was studied, leading 
to fully universal results, and conjectures for the correlation exponents. 
Some of these results were confronted to numerical simulations of an interface in $d=1$.
In a second part we studied truly dynamical correlations, between the velocities in the two avalanches.
We obtained the correlations of the total velocities and of the avalanche durations both for homogeneous
driving and for prescribed positions of the seeds. 
These correlations admit a fully universal massless limit which
we studied, leading to further conjectures for correlation exponents. 
Finally, we calculated the correlation between the shapes of two avalanches.
These were found to be subdominant for small avalanches, but non
zero for larger ones. It would be quite useful to probe these correlations further in
numerical simulations and in experiments to test the theoretical predictions. These tests
should allow to distinguish the various universality classes for avalanches.\\

Let us close by indicating an interesting direction for further work. Here we have 
shown that the correlations between avalanches separated by $W$ in the direction of
motion is proportional (to leading order) to $\Delta''(W)$, where $\Delta(w)$ is
the renormalized correlator of the pinning force. At the depinning fixed point
this quantity is {\it negative}, leading to anti-correlations. This result is valid at strictly
zero temperature. On the other hand, avalanches at very low but finite temperature were
studied recently in numerical simulations \cite{CreepAvalanchesFerrero}.
There "events" where the interface moves forward without returning,
similar to avalanches, 
were observed. These occur at scales below and around the 
so-called thermal activation nucleus scale (also called $L_{\rm opt}$).
These successive events tend to cluster in the same spatial region and 
are observed to be very strongly positively correlated
(reminiscent of the propagation of a forest fire). At larger scale, they 
appear to organize into clusters, which behave more like conventional
depinning avalanches. On the other hand, the FRG theory of creep,
as obtained in \cite{CreepChauve}, predicts a similar crossover in
scales from the creep to depinning regimes. It is well known that
in the creep regime $\Delta''(W)$ is {\it very large and 
positive} within a "thermal boundary layer" for small $W$, corresponding to the
thermal nucleus scale. We claim that this is quite consistent with the 
observations in Ref. \cite{CreepAvalanchesFerrero} of a strong 
positive correlation between the events. Obtaining a
detailed theory is more challenging, since it requires a precise
and operational definition of these events (as we did here for
the zero temperature avalanches). However we believe that our
result should provide the main guiding idea in that direction.

\section*{Acknowledgements}
We thank C. LePriol, L. Ponson and A. Rosso for interesting discussions.
We acknowledge 
support from ANR grant ANR-17-CE30-0027-01 RaMaTraF. TT's research was supported by a postdoctoral grant from the Research Foundation, Flanders (FWO).

\appendix

\section{Restoring units}
\label{app:restore} 

In this Appendix we give useful information on how to restore the dimensionfull units in the formula
for the problem with a mass $m>0$. To check units (and restore them) one must convert all quantities in units of 
$m, S_m,\tau_m$ which are the natural units. The conversion goes as follows
\bea
&& [x],[L]= m^{-1} \quad , \quad [w(x,t)],[u(x,t)] = S_m m^d \quad , \quad [t]=\tau_m
\quad , \quad [\dot u(x,t)] = S_m m^d/\tau_m \quad , \quad [\dot u^{\rm tot}]=v_m:= S_m/\tau_m \nn \\
&& [\Delta(w)]=m^{4-d} [w^2]= S_m^2 m^{4+d} \quad , \quad 
[\Delta'(w)] = S_m m^4 \quad , \quad [\Delta''(w)]=m^{4-d} 
\eea
As dimensional relations these are exact (i.e. up to dimensionless
prefactors) in any dimension. Note that the relation $-\Delta'(0)= S_m m^4$ is exact.
Let us give more details.

For kicks, source and response field: One has $\dot w(x,t)=\delta w(x) \delta(t)$, with for a uniform kick $\delta w(x)=\delta w$, then $[\delta w(x)]=[\delta w]=S_m m^d$. For the source, $[\lambda(x,t)]=1/S_m$, with the same unit for $\lambda(x,t)=\lambda$, conjugated to $S$. For the response field $[\tilde u(x,t)]=1/(S_m m^2)$. 

For local sizes: One has $[S(x)] = S_m m^d$ and $\lambda(x,t)=\lambda \delta^d(x)$ implies $[\lambda]=1/(m^d S_m)$.

For densities: one must distinguish densities for different driving, and for different observables, which have all different dimensions. The density w.r.t. a uniform driving is $[\rho(S)] = L^d/S_m^2$, $[\rho(\dot u^{tot})]= L^d/(S_m v_m)$.
The joint densities are $[\rho^c_W(S_1,S_2)] = [\rho_W(S_1,S_2)]= [\rho(S_1) \rho(S_2)]$.
The density w.r.t. uniform driving of local size is $[\rho(S(x))] = 1/(S_m^2 m^{2d})$
and $[\rho_W(S_1(x_1),S_2(x_2))] = 1/(S_m^4 m^{4d})$.
The densities with fixed kick positions have dimension $[\rho^x(S)] = 1/S^2_m$,  
$[\rho^{c,x_1,x_2}_W(S_1,S_2)] = [\rho^{c,x}_W(S_1,S_2)]= 1/S^4_m$,
and in Fourier space 
$[\rho^{c,q}_W(S_1,S_2)] = 1/(m^d S^4_m)$. Similarly one has 
$[\rho^x(\dot u)]= 1/(S_m v_m)$ 
$[\rho^{c,x_1,x_2}_W(\dot u_1,\dot u_2)]= 1/(S_m^2 v_m^2)$
$[\rho^{c,q}_W(\dot u_1,\dot u_2)]= 1/(m^d S_m^2 v_m^2)$.

All the above assumes that one converts $L=1/m$. There are also some rules for
how $L$ appears in the formula. For instance densities with uniform driving are
$\rho(S) \sim L^d$,
$\rho_W(S_1,S_2) \sim L^{2d}$, while connected densities are
$\rho^c_W(S_1,S_2) \sim L^{d}$. The densities with fixed seed positions
are all $O(1)$.

In a second stage one can make explicit the dependence in $m$, and
introduce the roughness and dynamical exponents, i.e. write $S_m = A_S m^{-(d+\zeta)}$ 
and $\tau_m = A_\tau m^{-z}$, and write $\Delta(w) = m^{\epsilon-2 \zeta} 
\hat \Delta(m^\zeta w)$. 

Finally, what we call the "massless dimensionless units" are such that $\sigma=\eta=1$.

\section{Avalanche decomposition}
\label{app:avalanches} 

Let us justify further the formula \eqref{expansion}. The avalanche picture is the following.
In the limit of very slow driving, one can assume that the part of the
velocity field $\dot u \equiv \dot u(x,t)$ which is $O(1)$ can be decomposed
in a sum over discrete events called avalanches, schematically 
\be \label{sum}
\dot u = \sum_\alpha \dot u^{(\alpha)} 
\ee
Each $\dot u^{(\alpha)}$ is a random velocity field (inside one avalanche).
It is either non-zero (the avalanche has occured) or zero $\dot u^{(\alpha)}=0$
with finite probability (the avalanche has not occured). In practice
it means that the velocity is not $O(1)$, it can be non-zero 
but vanishes as the driving vanishes.

Now we can use the identity 
\be \label{identity}
e^{ \lambda \cdot \dot u} = \prod_\alpha e^{\lambda \cdot \dot u^{(\alpha)}  } 
= \sum_{n=0}^{+\infty} \sum_{1 < \alpha_1 < \alpha_2 < \dots \alpha_n} \prod_{j=1}^n 
(e^{\lambda \cdot \dot u^{(\alpha_j)}  } -1) 
\ee
If we want to think of each avalanche $\alpha$ to be triggered by a
small kick $\delta w_\alpha$ with probability proportional to $\delta w_\alpha$,
we can average \eqref{identity} and obtain
\be \label{identity2}
G[\lambda] = \overline{\prod_\alpha e^{\lambda \cdot \dot u^{(\alpha)}  } }
= \sum_{n=0}^{+\infty} \sum_{1 < \alpha_1 < \alpha_2 < \dots \alpha_n} 
\prod_{j=1}^n \delta w_{\alpha_j} \int \prod_{j=1}^n D[\dot u^{(\alpha_j)}] \, \rho_{\alpha_1,..\alpha_n}(\dot u^{(\alpha_1)}
,..,\dot u^{(\alpha_n)})
\prod_{j=1}^n (e^{\lambda \cdot \dot u^{(\alpha_j)}  } -1) 
\ee
where 
\bea
(\prod_{j=1}^n \delta w_{\alpha_j}) \rho_{\alpha_1,..\alpha_n}(\dot u^{(\alpha_1)}
,..,\dot u^{(\alpha_n)})
\eea 
is the probability that the avalanches $\alpha_1,.. \alpha_n$ 
have occured, and $\rho_{\alpha_1,..\alpha_n}(\dot u^{(\alpha_1)}
,..,\dot u^{(\alpha_n)})$ is the associated joint density for the avalanche velocities
to take values $\dot u^{(\alpha_1)},.. ,\dot u^{(\alpha_n)}$. In the BFM the avalanches are independent, and 
these densities are just products $\rho_{\alpha_1,..\alpha_n}(\dot u^{(\alpha_1)}
,..,\dot u^{(\alpha_n)}) = \prod_{j=1}^n \rho_{\alpha_j}(\dot u^{(\alpha_j)})$ and one
obtains 
$ e^{ \lambda \cdot \dot u} = e^{\sum_\alpha \delta w_\alpha 
\int D[\dot u^{(\alpha)}] \rho_\alpha(\dot u^{(\alpha)}) (e^{\lambda \cdot \dot u^{(\alpha)}  } -1) }$.

If we consider a source $\lambda$ which is non-zero only for two specific avalanches,
we see that we obtain the formula \eqref{expansion} since all other terms vanish. 
There is a small subtelty however concerning
e.g. the terms $\delta w_1^2, \delta w_2^2$ in \eqref{expansion} (not of interest there). 
The above picture is correct for
distinct kicks $\delta w_\alpha$. There are additional terms in $G[\lambda]$ containing 
powers of $\delta w_\alpha^p$ with $p>1$. Those are obtained by a small modification,
namely there can be in the sum \eqref{sum} $p_\alpha$ replica of the same avalanche
(i.e. the kick $\delta w_\alpha$ can trigger $p$ avalanches).
In the BFM the $p_\alpha$ are distributed according to the Poisson distribution. Here we are not
interested in these terms and we can use $p_\alpha=1$. Hence the above picture is sufficient.\\

One can now compare with the expansion of \eqref{action2} in powers of $\dot w_{xt}$, namely
\bea \label{expansionw}
G[\lambda] = 1 + m^2 \int_{xt} \dot w_{xt} \langle \tilde{u}_{xt} \rangle_{S_\lambda} 
+ \frac{1}{2} m^4 \int_{xt,y't'} \dot w_{xt} \dot w_{yt'}  \langle \tilde{u}_{xt} 
 \tilde{u}_{yt'}  \rangle_{S_\lambda} + \dots
\eea 
where the brackets denote expectations of the response fields in the 
theory $S_\lambda = S - \int_{xt} \lambda_{xt} \dot{u}_{xt}$ (normalized since
$G[\lambda]=1$ when $\dot w_{xt}=0$). Choosing $\dot w_{xt}$ to
be a series of kicks at well separed times (much larger than the typical 
duration of an avalanche) $\dot w_{xt}=\sum_\alpha 
\delta w_\alpha(x) \delta(t-t_\alpha)$, and inserting in \eqref{expansionw}
we obtain an expansion similar to \eqref{identity2}, and we can identify, e.g.
\bea
&& \int D[\dot u^{(\alpha)}]  \rho^x_{\alpha}(\dot u^{(\alpha)}) (e^{\lambda \cdot \dot u^{(\alpha)}  } -1) 
= m^2 \langle \tilde{u}_{x,t_\alpha} \rangle_{S_\lambda}  \\
&& \int D[\dot u^{(\alpha)}] D[\dot u^{(\beta)}]   \rho^{x,y}_{\alpha,\beta}(\dot u^{(\alpha)},\dot u^{(\beta)})(e^{\lambda \cdot \dot u^{(\alpha)}  } -1) (e^{\lambda \cdot \dot u^{(\beta)}  } -1) 
= m^4 \langle \tilde{u}_{x,t_\alpha} \tilde{u}_{y,t_\beta} \rangle_{S_\lambda} \label{Slambda} 
\eea 
and so on ($\alpha \neq \beta$ in the second relation). In the BFM 
$ \langle \tilde{u}_{x,t} \rangle_{S_\lambda} = \tilde{u}^\lambda_{x,t}$,
$ \langle \tilde{u}_{x,t} \tilde{u}_{y,t'} \rangle_{S_\lambda} = \tilde{u}^\lambda_{x,t} \tilde{u}^\lambda_{y,t'}$
and so on in terms of the solution of the instanton equation, 
and the densities factorize. The calculations performed in the main text amount
to calculate these expectation values beyond the BFM. As said above the two times $t,t'$ in \eqref{Slambda} 
are very far away and chosen to belong to different avalanches. The response field
correlation in \eqref{Slambda} when $t,t'$ are distant of order $\tau_m$ instead
allows to study overlapping avalanches, which goes beyond the present study.

\section{Derivation of the action}
\label{app:derivation} 

In this Appendix we justify our main result \eqref{GenerFT1}-\eqref{ActionTotatle}-\eqref{ActionBFM} about the simplified field theory which allows to compute correlations between several avalanches at order $O(\epsilon)$ in the depinning dynamics of elastic interfaces. 
To this aim, it is easier to consider the first protocol (see Sec.~\ref{subsec:Model}) where the interface is driven at a constant velocity $v  \to 0^+$. For compactness we denote the space and time dependence in
subscript i.e. $u(x,t) \equiv u_{xt}$, $\lambda(x,t) \equiv \lambda_{xt}$ and so on.\\

%We construct the field theory for correlations between avalanches in that context as the field theory has been mostly studied in that protocol, 
%but the results presented in the text is just a reformulation of what is derived here.
%

Our starting point is that the MSR action for the velocity theory is exactly given by $S[\tilde u , \dot u]=S_0[\tilde u , \dot u]+S_{\rm dis}[\tilde u , \dot u] - m^2   \int_{xt} \dot w_{xt} \tilde{u}_{xt}$ with (see Eqs.~(301)-(303) in 
\cite{LeDoussalWiese2012a})
\bea
&& S_0[\tilde u , \dot u]= \int_{xt} \tilde{u}_{xt} (\eta \partial_t - \nabla_x^2 + m^2) \dot{u}_{xt}   \nn \\
&&  S_{\rm dis}[\tilde u , \dot u] = - \sigma \int_{x t} \tilde{u}_{xt}^2 \dot{u}_{xt}+ \frac{1}{2} \int_{x t t'}  \tilde{u}_{xt} \tilde{u}_{xt'}\dot{u}_{xt}  \dot{u}_{xt'} \Delta_{\rm reg}''( u_{xt} - u_{xt'}) \, .
\eea
where $\sigma = - \Delta'(0^+)$ and $\Delta_{\rm reg}''(u)$ is the regularized version of the renormalized disorder correlator $\Delta(u)$ that is smooth at $0$ and defined by $\Delta_{\rm reg}(u) = \Delta(u) + \sigma |u|$.\\

This action allows to calculate observables of the velocity field as, for any source $\lambda_{xt}$,
\bea \label{action2} 
G[\lambda] = \overline{ e^{\int_{xt} \lambda_{xt} \dot{u}_{xt}}} = \int {\cal D}[\tilde u , \dot{u}] e^{\int_{xt} \lambda_{xt} \dot{u}_{xt} - S_0 [\tilde u , \dot u] - S_{\rm dis}[\tilde u , \dot u] + m^2   \int_{xt} \dot w_{xt} \tilde{u}_{xt} } \, .
\eea

Let us consider the slow uniform driving $\dot w_{xt}=v \to 0^+$.
Being interested in correlations between different avalanches, we can consider a source of the form
\bea \label{app:observable}
\lambda_{xt} = \lambda_1(x,t) \theta(t) \theta({\sf T}_1 - t) + \lambda_2(x,t) \theta(t-W/v) \theta(W/v+{\sf T}_2 - t) \, .
\eea
That is, as source that is active in two different time windows $[0,{\sf T}_1]$ and $[W/v,W/v+ {\sf T}_2]$ and probes the result of two avalanches  eventually occuring at times $ t \in [0,{\sf T}_1]$ and $t \in [W/v,W/v+ {\sf T}_2]$. Taking first the limit $v \to 0$ with ${\sf T}_1$ and ${\sf T}_2$ fixed, it is clear that if the interface moves during both time windows, this is due to {\it different avalanches} since the duration of an avalanche is $O(v^0)$.\\

It is a crucial point that near the critical dimension one has $u_{xt} = vt + O(\epsilon)$, while $\Delta_{{\rm reg}}''$ is also uniformly $O(\epsilon)$. That means that, at order $O(\epsilon)$, we can replace in the above action $\Delta_{\rm reg}''( u_{xt} - u_{xt'})  \to \Delta_{\rm reg}''( v (t-t'))  + O(\epsilon^2) $. We now rescale the fields $\tilde{u}$ and $\dot{u}$ by introducing the characteristic scales of the avalanche motion: we rescale $t \to \tau_m t$, $x \to m^{-1} x$ and $u \to m^d S_m u$ with $\tau_m = \eta/m^2$ and $S_m = \sigma/m^4$. That leads to a rescaling of the fields as $\dot{u} \to v_m \dot{u}$ and $\tilde{u} \to \frac{1}{m^2 S_m} \tilde{u}$ with $v_m =  m^d S_m /\tau_m$. We also use that for $m$ close to $0$ the renormalized disorder correlator $\Delta$ takes a scaling form $\Delta(u) =  m^{\epsilon -  2 \zeta} \hat{\Delta}(m^\zeta u)$ with $\hat{\Delta}$ a function that converges to a FRG fixed point uniformly of order $O(\epsilon)$
in the $m \to 0$ limit and $\zeta$ the roughness exponent of the interface. Rescaling finally the source field as $\lambda_{xt} \to \frac{1}{S_m} \lambda_{xt}$ and the driving velocity as $v \to v_m v$ we can decompose the action for these rescaled variables between the tree action ($O(\epsilon^0)$) and one-loop corrections ($O(\epsilon)$)
\be
G[\lambda] = \overline{ e^{\int_{xt} \lambda_{xt} \dot{u}_{xt}}} = \int {\cal D}[\tilde u , \dot{u}] e^{\int_{xt} \lambda_{xt} \dot{u}_{xt} - S_{\rm tree} [\tilde u , \dot u] -  \delta S_{\rm loop}[\tilde u , \dot u] + v  \int_{xt} \tilde{u}_{xt} }
\ee
with the tree action corresponding to the rescaled version of the BFM action
\be
S_{\rm tree} [\tilde u , \dot u]  = \int_{xt} \tilde{u}_{xt} (\partial_t - \nabla_x^2 + 1) \dot{u}_{xt}  -  \int_{x t} \tilde{u}_{xt}^2 \dot{u}_{xt} \sim O(1) 
\ee
and the $O(\epsilon)$ corrections
\be
\delta S_{\rm loop}[\tilde u , \dot u]  = \frac{1}{2} \int_{x t t'}  \tilde{u}_{xt} \tilde{u}_{xt'}\dot{u}_{xt}  \dot{u}_{xt'} \hat{\Delta}_{\rm reg}'' \left( m^\zeta v_m v  \tau_m (t-t') \right) \, \sim O(\epsilon).
\ee

On the other hand, expanding $G$ at order $v^2$ we obtain
\bea \label{app:GlambdaExpansion}
G[\lambda] =&&  1 + v  \int {\cal D}[\tilde u , \dot{u}]  \int_{xt} \tilde{u}_{xt}  e^{\int_{xt} \lambda_{xt} \dot{u}_{xt} - S_{\rm tree} [\tilde u , \dot u] - \delta S_{\rm loop}[\tilde u , \dot u] } \nn \\
&&  + \frac{v^2}{2} \int {\cal D}[\tilde u , \dot{u}]  \int_{xt} \tilde{u}_{xt}  \int_{x't} \tilde{u}_{x't'}  e^{\int_{xt} \lambda_{xt} \dot{u}_{xt}- S_{\rm tree} [\tilde u , \dot u] - \delta S_{\rm loop}[\tilde u , \dot u] }  + O(v^3) \, .
\eea
As it has been already discussed numerous times in that context, e.g.
\cite{LeDoussalWiese2012a,ThieryShape}, each response field $\tilde{u}_{xt}$ present in front of the dynamical path weight in \eqref{app:GlambdaExpansion} generates the contribution to the observable $G[\lambda]$ that comes from an avalanche starting at time $t$ at position $x$. The first line in \eqref{app:GlambdaExpansion} thus corresponds to the contribution from single avalanches (a single avalanche occured with probability of order $v$), while the second line corresponds to the contribution from two avalanches (two avalanches occured with probability of order $v^2$), and contain the correlations between the two avalanches. Let us now think a bit diagrammatically about the calculation of an observable like \eqref{app:observable} at order $O(\epsilon)$. This can be performed by an expansion
in $\lambda_{xt}$ and $\delta S_{\rm loop}$ of the $O(v^2)$ term in \eqref{app:GlambdaExpansion}
and computing the resulting correlation functions involving
the fields $\tilde{u}_{xt}$ and $\dot{u}_{xt}$ within the $S_{\rm tree}$ action. 
%Each field $\tilde{u}_{xt}$ can first be contracted with a field $\dot{u}_{xt}$ living inside the first or second time window. 
Since the tree action can only connect fields at times differing by a time scale at most of
order $O(1) \ll  1/v$ and that the $O(\epsilon)$ interaction vertex $\delta S_{\rm loop}[\tilde{u} , \dot{u}]$ can only be used once at this order, it is clear that we can only get diagrams of two types at order $O(\epsilon)$. The first type are diagrams where the $\delta S_{\rm loop}$ term was used to contract fields in the same time-window. This leads to diagrams where fields on the two time-windows are disconnected. These do not participate to the correlations between the two avalanches at $w=0$ and $w=W$. For these diagrams we can replace in the limit $v \to 0$, $\hat{\Delta}_{\rm reg}'' \left( m^\zeta v_m v  \tau_m (t-t') \right) \to \hat{\Delta}_{\rm reg}''(0)$ in $\delta S_{\rm loop}$. In the second type of diagrams 
the $\delta S_{\rm loop}$ term is used to contract fields in different time windows. These are the only diagrams that contribute to correlations between avalanches at $w=0$ and $w=W$. For these diagrams we can replace in the limit $v \to 0$, $\hat{\Delta}_{\rm reg}'' \left( m^\zeta v_m v  \tau_m (t-t') \right) \to \hat{\Delta}_{\rm reg}''(m^\zeta W)$ in $\delta S_{\rm loop}$.\\

In the above discussion, the first type of diagrams contains the diagrams that lead to the $O(\epsilon)$ corrections to the single avalanche statistics as studied in \cite{LeDoussalWiese2012a,ThieryShape}. The second type of diagrams on the other hand generates the correlations between avalanches as studied in this paper. Since fields living inside the two different time windows (avalanches) can only be connected once by the $\hat \Delta''(W)$ interaction vertex we can formally introduce two different copies of the fields, one for each time window, the two copies being only connected by the $\hat \Delta''(W)$ interaction vertex. The contribution to $G[\lambda]$ coming from avalanches starting at position $x_1$ at time $t=0$ and at position $x_2$ at time $t=W/v$ can then be targeted by restricting the response fields outside the exponential in the second line of \eqref{app:GlambdaExpansion} to $(x_1,t_1)$ and $(x_2,t_2)$. Once this has been done one can send the time window ${\sf T}_1$ and ${\sf T}_2$ to infinity to ensure that the avalanche terminates inside the time window with probability $1$ (the order ${\sf T}_i \ll 1/v$ holds since the limit $v \to 0$ has already been taken). Going back to the original units of the problem, one then sees that \eqref{app:GlambdaExpansion} leads to 
\eqref{rhoWform} and more generally we obtain the formulation of the theory presented in the text in
Eqs. \eqref{GenerFT1}-\eqref{ActionTotatle}-\eqref{ActionBFM}.

\section{Formula for the shape correlation at fixed seed positions} 
\label{app:formula}

We give here the formula for arbitrary $q$ in \eqref{shapeu} 
We have
\bea
&& \partial_{\lambda_i}|_{\lambda_i=0} \int_0^{t'_i} dt' G_i(q,0,t') \tilde u^0_i(t')  = 
 \int_0^{t_i} dt' e^{- (q^2+1) t' } \frac{e^{t'-t_i} (e^{t'_i}-e^{t_i})^2}{(e^{t_i'} -1)^2}   \\
 && -2  \int_0^{t_i} dt' e^{- (q^2+1) t' } 
 \frac{e^{t'- t_i} (e^{t'_i} - e^{t_i})^2 (e^{t'}-1)}{(e^{t_i'}-1)^3}  - 2 \int_{t_i}^{t'_i} dt' e^{- (q^2+1) t' } \frac{e^{t'-t_i} (e^{t_i'}-e^{t'})}{(e^{t_i'} - 1)^3}  
  (e^{t_i} - 1) (e^{t'_i}-e^{t_i}) \\
% \eea 
%This gives
%\bea
%&& \int_0^{t'_i} dt' G_i(q,0,t') \tilde u^0_i(t')  = \frac{e^{-t_i} (e^{t'_i}-e^{t_i})^2}{(e^{t_i'} -1)^2}  
% \int_0^{t_i} dt' e^{- q^2 t' }   \\
% && -2   \frac{e^{- t_i} (e^{t'_i} - e^{t_i})^2}{(e^{t_i'}-1)^3}  
% \int_0^{t_i} dt' e^{- q^2 t' } (e^{t'}-1) \\
%&& - 2 \frac{e^{-t_i} (e^{t_i} - 1) (e^{t'_i}-e^{t_i}) }{(e^{t_i'} - 1)^3}  
%   \int_{t_i}^{t'_i} dt' e^{- q^2 t' } (e^{t_i'}-e^{t'})
% \eea 
%\bea
&&  = 
\frac{e^{-t_i} (e^{t'_i}-e^{t_i})}{(e^{t_i'} -1)^2}   \bigg( (e^{t'_i}-e^{t_i}) \int_0^{t_i} dt' e^{- q^2 t' }  -2   \frac{ (e^{t'_i} - e^{t_i})}{(e^{t_i'}-1)}  
 \int_0^{t_i} dt' e^{- q^2 t' } (e^{t'}-1)  - 2 \frac{ (e^{t_i} - 1)  }{(e^{t_i'} - 1)}  
   \int_{t_i}^{t'_i} dt' e^{- q^2 t' } (e^{t_i'}-e^{t'}) \bigg) \nonumber \\
&& = \frac{e^{-t_i} \left(e^{t'_i}-e^{t_i}\right)
   \left(\left(1-e^{-q^2 t_i}\right)
   \left(e^{t'_i}-e^{t_i}\right)+\frac{2 \left(e^{q^2
   \left(-t_i\right)+t'_i+t_i}+e^{t'_i-q^2
   t'_i}-e^{q^2
   \left(-t'_i\right)+t'_i+t_i}-e^{t_i-q^2
   t_i}-e^{t'_i}+e^{t_i}\right)}{\left(q^2-1\right)
   \left(e^{t'_i}-1\right)}\right)}{q^2
   \left(e^{t'_i}-1\right){}^2}
 \eea 
from which we obtain the result by a simple derivative
\bea \label{resapp}
r_{q,t_i}(T_i) =   \partial_{t'_i}|_{t'_i=T_i} \partial_{\lambda_i}|_{\lambda_i=0} \int_0^{t'_i} dt'  G_i(q,0,t') \tilde u^0_i(t')
\eea

%In the limit $q=0$ one finds
%\bea
%\frac{1}{2} \sinh \left(\frac{1}{2}
%   \left(t_i-t'_i\right)\right) \left(t_i \sinh
%   \left(\frac{1}{2} \left(t_i-2
%   t'_i\right)\right)-\left(t_i-2 t'_i\right) \sinh
%   \left(\frac{t_i}{2}\right)\right)
%   \text{csch}^3\left(\frac{t'_i}{2}\right)
%\eea 
%
%Taking the derivative one gets
%\bea
%-\frac{1}{4} \text{csch}^4\left(\frac{T}{2}\right)
%   (\sinh (t-T)-(t-2 T) \cosh (t)+(t+T) \cosh
%   (t-T)-\sinh (t)-2 T+\sinh (T)+T (-\cosh (T)))
%\eea 

\end{widetext}

\end{document}